\documentclass[acmtog,nonacm]{acmart}
\acmSubmissionID{1234}

\usepackage{booktabs} 

\usepackage{svg}
\usepackage{bm}
\usepackage{amsmath}
\usepackage{subcaption}
\usepackage{multirow}
\usepackage{pifont}
\usepackage{rotating}
\usepackage{paralist}
\usepackage{placeins}
\usepackage[ruled]{algorithm2e} 

\SetAlFnt{\small}
\SetAlCapFnt{\small}
\SetAlCapNameFnt{\small}
\SetAlCapHSkip{0pt}

\newcommand{\xmark}{\ding{55}}
\newcommand{\cmark}{\ding{51}}
\acmJournal{TOG}

\newcommand{\ready}[1]{}

\newcommand{\methodname}{VolTeMorph}

\makeatletter
\DeclareRobustCommand\onedot{\futurelet\@let@token\@onedot}
\def\@onedot{\ifx\@let@token.\else.\null\fi\xspace}
\def\eg{\emph{e.g}\onedot} 
\def\etal{\emph{et al}\onedot}

\begin{document}

\title{\textbf{VolTeMorph}: Realtime, Controllable and Generalisable Animation of Volumetric Representations}

\author{Stephan J. Garbin}
\authornote{Denotes equal contribution.}
\author{Marek Kowalski}
\authornotemark[1]
\author{Virginia Estellers}
\authornotemark[1]
\author{Stanislaw Szymanowicz}
\authornotemark[1]
\author{Shideh Rezaeifar}
\author{Jingjing Shen}
\author{Matthew Johnson}
\author{Julien Valentin}
\affiliation{\institution{Microsoft}}

\authorsaddresses{}


%
%


\begin{teaserfigure}
\centering
\includegraphics[width=7.0in]{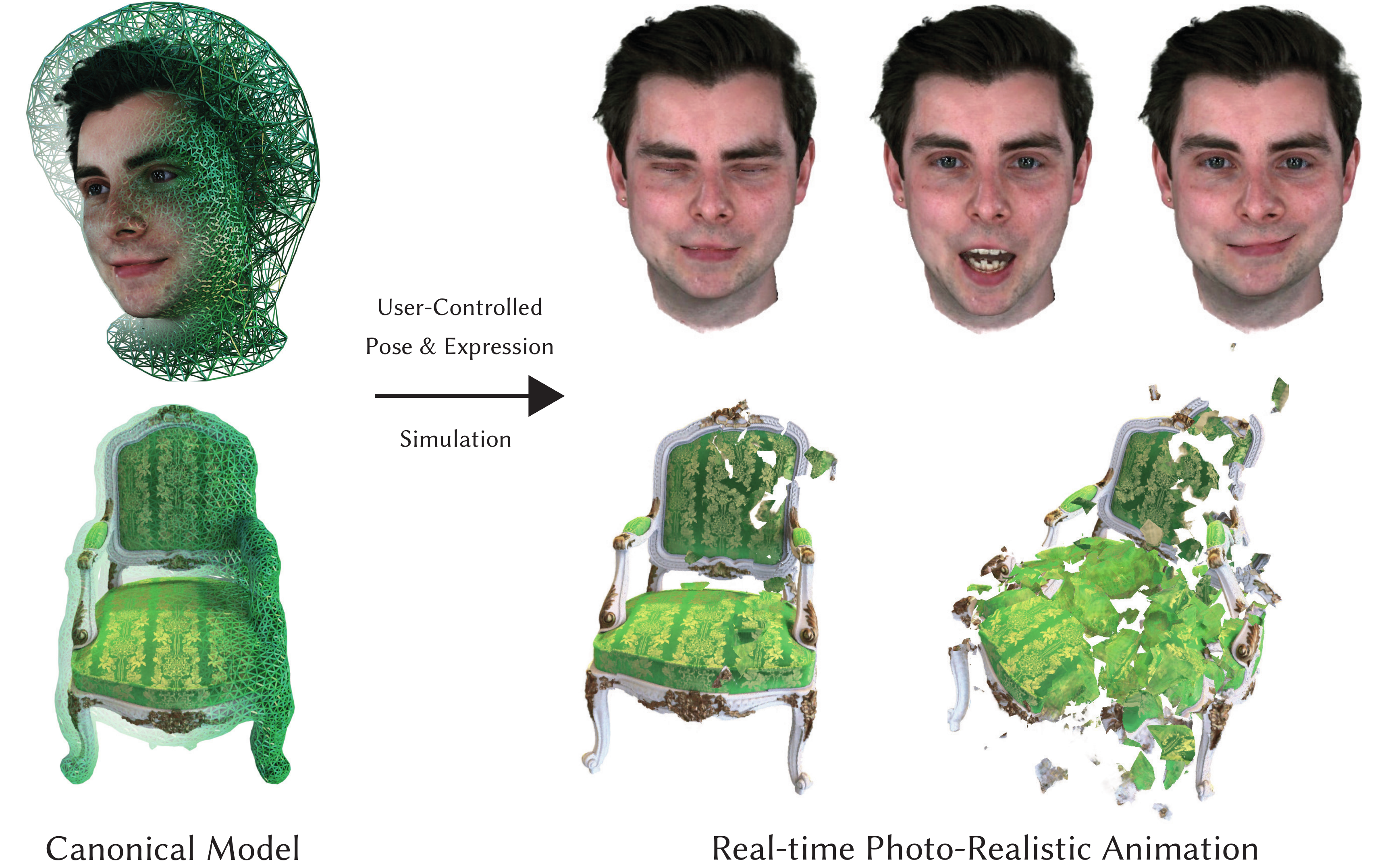}
\caption{\ready{} We propose a method to deform static multi-view volumetric models, such as NeRF, in real-time using blendshape or physics-driven animation. This allows us to create dynamic scenes from static captures in an interpretable, artistically controllable way. Top: a face tracker controls an avatar which shows expressions unseen during training. Bottom: object shattering controlled by physics-based simulation.}
\label{fig:teaserfigure}
\end{teaserfigure}
\begin{abstract}
\ready{}
The recent increase in popularity of volumetric representations for scene reconstruction and novel view synthesis has put renewed focus on animating volumetric content at high visual quality and in real-time. While implicit deformation methods based on learned functions can produce impressive results, they are `black boxes' to artists and content creators, they require large amounts of training data to generalise meaningfully, and they do not produce realistic extrapolations outside the training data. 
In this work we solve these issues by introducing a volume deformation method which is real-time, easy to edit with off-the-shelf software and can extrapolate convincingly.
To demonstrate the versatility of our method, we apply it in two scenarios: physics-based object deformation and telepresence where avatars are controlled using blendshapes.
We also perform thorough experiments showing that our method compares favourably to both volumetric approaches combined with implicit deformation and methods based on mesh deformation.


\end{abstract}
\maketitle

\section{Introduction}
\ready{}
Neural Radiance Fields (NeRF)~\cite{nerf} is a method for generating 3D content from images taken with commodity cameras that has prompted a major change in the field. The main limitations of NeRF are its rendering speed and being constrained to static scenes. Rendering speed has been successfully addressed by multiple follow-up works \eg~\cite{SnergBaked,garbin2021fastnerf,plenoctrees}, but the constraint to static scenes remains an open challenge to deploying this technology in interactive applications. In this work we present a technique for generating photo-realistic dynamic content in a scalable way. While we use a neural radiance field for its photo-realism, our technique is compatible with most volumetric representations built on neural, explicit or even hash-coded voxel representations~\cite{plenoxels,instantNGP}.

We focus on four major aspects of scalability. \textbf{Scalable rendering}: ability to render in real-time on commodity graphics hardware. \textbf{Scalable content}: support for different types of content from generic objects to personalized human avatars. \textbf{Scalable enrolment}: capacity of training with minimal amount of data and ability to generalize to unseen dynamics. \textbf{Scalable control}: compatibility with existing control mechanisms for animation, for instance using physics-based simulation for virtual objects or blendshape-based face tracking for virtual heads.

Scalable control is critical to overcome the limitations of existing methods for animating NeRF-based scene representations, which either cannot deform beyond motion seen during training or only support specific objects and motions (see Table~\ref{tab:scalability_comparison}). For instance, \cite{nerfies,Gafni_2021_CVPR} are capable of high quality interpolation between the training samples but are not designed to be robust to extrapolations, while \cite{avaps,neural_head_avatars} are capable of extrapolating outside of the training data but only support human heads. We propose a technique for \textbf{Vol}umetric \textbf{Te}trahedral \textbf{Morph}ing, VolTeMorph, that allows for realistic-looking extrapolations and equally supports generic objects and humans.

\begin{table}
    \centering
    \begin{tabular}{c|c|c|c|c}
        \multirow{2}{*}{Method} & \multicolumn{4}{c}{Axis of Scalability} \\
        {} & Enrol & Control & Render & Content \\
        \hline
        AVAPS~\cite{avaps} & \cmark & \cmark & \cmark & \xmark \\
        NHA~\cite{neural_head_avatars} & \cmark & \cmark & \cmark  & \xmark \\
        Nerfies~\cite{nerfies} & \xmark & \cmark & \xmark & \cmark \\
        NerFace~\cite{Gafni_2021_CVPR} & \xmark & \cmark & \xmark & \xmark \\
        Ours & \cmark & \cmark & \cmark  & \cmark
 \\
    \end{tabular}
    \caption{
    Comparison of our method to other popular solutions for animating NeRF-style models. The evaluated factor is scalability in terms of: enrolment, provided control over generated content, speed of rendering as well as the ability to adapt to different types of content.}
    \label{tab:scalability_comparison}
\end{table}

At the core of \methodname{} lie tetrahedral volumetric elements that compactly capture dynamic contents while allowing us to preserve the visual quality of NeRF. The resulting volumetric mesh can be intuitively deformed and controlled by artists, physics-based simulation, or traditional animation techniques like blendshapes. To demonstrate the versatility of \methodname{}, we perform experiments with two very different types of motion and objects.

In our first application we use physics-based simulation to control the deformation of a static object undergoing complex topological changes and render photo-realistic images of the process for every step of the simulation. This application is designed to show the representation power of our deformation model and the ability to render images from physical deformations difficult to capture with a camera. For that reason this series of experiments is conducted on synthetic data.

In our second application, we demonstrate photo-realistic animations of human head avatars in real-time with a blendshape-based face tracker similar to~\cite{wft}. The avatars are trained with 30 images of the subject taken from different viewpoints at the same instant. Thus, for each avatar the method has only seen a single face expression and pose. To animate the head avatars we use the control parameters of the parametric 3DMM face model that we extend from a surface mesh to the volume around it. We call the resulting parametric volumetric face model Vol3DMM. Building on the parametric face model allows us to generalize to face expressions and poses unseen at training and to drive the animation using face trackers built for real-time control. A key benefit of our method over traditional 3DMM face models is that hair, accessories and other elements not modelled by the 3DMM are captured by the volumetric geometry. While we demonstrate results on faces, we believe that the proposed technique to generalize 3DMM models to volumetric representations could be applied to full bodies \cite{SMPL:2015}.

In summary, our main contributions are:
\begin{itemize}
    \item \methodname{}, a method to animate volumetric scene representations with well-understood techniques such as blendshape animation and physics-based simulation.
    \item Vol3DMM, a volumetric extension to the traditional 3DMM allowing our system to be integrated seamlessly with existing tools for face  tracking.
    \item A scalable, real-time system for rendering photorealistic avatars of human faces built on \methodname{} and Vol3DMM.
\end{itemize}

\section{Related work}
\label{section:related_work}
\ready{}

Despite our method being a generic approach to deforming volumes, we demonstrate the animation of static face scans as our main area of application. Representation and animation of faces also serves as a relevant way to structure the literature around mesh-based and volumetric neural rendering, where faces are often chosen as a particularly difficult and sought after use-case. We do not discuss purely or 3D-assisted neural rendering methods such as~\cite{deferredNeuralRendering, kim2018deepVideoPortraits}, and refer the interested reader to~\cite{zollhoefer2018facestar} (Section $10$) for a more in-depth review.

\subsection{Mesh-Based Neural Rendering}
Codec Avatars~\cite{originalCodecAvatarsPaper} demonstrated that meshes in combination with textures generated by a Variational Autoencoder (VAE) conditioned on expression and view direction could be employed to generate highly realistic avatars in real-time given large amounts of multi-view training data.

Relying on only one camera to reduce the data overhead of such methods, the authors in Neural Head Avatars~\cite{neural_head_avatars} model a deforming head by first fitting a parametric model (FLAME)~\cite{flame_citation}, and subsequently using optimised pose and expression parameters to generate displacements and textures with learned functions. This assumes a fixed view direction. I~M~Avatar is another recent example~\cite{zheng2022imavatar} that learns blendshape correctives and textures directly from monocular video data.

In contrast to these approaches that use several hundred frames, we train on a maximum of two frames of a multi-view capture only. We also use a volumetric equivalent of a 3DMM, such as FLAME, which allows us to capture hair with greater fidelity, faithfully representing transparency  for composites over background content.

\subsection{Volumetric Neural Rendering}
Given the limitations of mesh and texture based representations to model things such as vegetation, hair, clouds, or any other class of object inherently easier to represent volumetrically~\cite{Vicini2021NonExponential}, learned 3D representations have grown in popularity in recent years.

In Neural Volumes, Lombardi~\etal learn the parameters, i.e. colour and density, of a purely emissive volume on a dense grid to represent bounded scenes.

NeRF~\cite{nerf} showed that such a volumetric function can be compressed using a single Multi-Layer Perceptron (MLP) in combination with a frequency encoding that enables the MLP to learn complex functions in two and three dimensions. These encodings were previously proposed for different applications by~\cite{convSequence2Sequence, attentionIsAllYouNeed}, although a single-scale variant was used in computer graphics by~\cite{neuralImportanceSampling}. The notion of frequency encodings was subsequently generalised in~\cite{fourierFeatNetworks}.

Since NeRF's initial breakthrough, many methods that mix together implicit functions and explicit voxel grids have emerged~\cite{garbin2021fastnerf, SnergBaked, plenoctrees, directVoxelOptimisation, neuralSparseVoxelFields}. We note that all of these are compatible with \methodname. This is because whether represented implicitly or explicitly, these representations are fundamentally volumetric functions in $\mathbb{R}^3$, and can be animated by a displacement field in that space.

\subsubsection{Volumetric Deformation Using Learned Functions}
In Neural Volumes, Lombardi~\etal learn a model that generates inverse warp fields to map voxels back to a canonical space~\cite{neuralVolumes}. Neural Body is a variation of this concept, where a latent code volume is produced by diffusing features anchored to a SMPL~\cite{SMPL:2015} model to a voxel grid, and subsequently feeding them through a decoder network~\cite{neuralBody2021}.

D-NeRF uses an implicit deformation model that maps sample positions back to a canonical space~\cite{dNeRF}. Two common issues with models that use (implicit) deformation functions like D-NeRF are (a) lack of generalisation to unseen deformations and (b) runtime overheads due to querying a deformation network MLP, potentially millions of times per frame for coordinate-based models.

Nevertheless, Nerfies~\cite{nerfies} and its extension~\cite{hyperNerf} use a similar concept as~\cite{dNeRF}, but account for changes in the observed scenes with a per-image latent code which enables modelling changes in colour in addition to shape. This is crucial to allow for modelling dynamic scenes with multiple temporal frames of training data. For example a person might blink or otherwise move their face between subsequent observations. The authors in Nerfies also propose a rigidity constraint on the implicit learned canonical mapping to prevent overfitting. The model by Gafni~\etal opts to use a single implicit network conditioned on a face tracking signal from a 3D morphable model as well as a per-image latent code to an implicit volumetric model of animated human faces from monocular video~\cite{Gafni_2021_CVPR}. An implicit deformation model is also employed in Non-rigid NeRF~\cite{nonrigidNeRF}, but augmented with a segmentation of the scene into rigid and deforming parts, as well as a non-divergence constraint of the deformation field. 
In contrast to these approaches, we focus on using as little temporal enrolment data as possible while still ensuring generalisation by exploiting prior knowledge encoded in our volumetric blendshape model in order to address the issue of bad generalisation. We circumvent the issue of performance by proposing a real-time algorithm for deformation that runs on commodity graphics hardware.

\subsubsection{Hybrid Volumetric Deformation}
The Mixture of Volumetric Primitives (MVP) model represents a return to an explicit volumetric representation. Instead of one dense grid, a convolutional decoder generates many small, rectangular volumetric primitives conditioned on an input signal representing view direction and other latent information~\cite{mixtureVolumetricPrimitives}. These primitives are anchored to a tracked face mesh in a learned way, and can be rendered efficiently. Using a large corpus of 3D captures, an approach built on a similar representation was recently shown to generalise given limited enrolment data captured using handheld devices~\cite{avaps}.

\subsection{Volumetric Deformation using Geometry}
In terms of using a surface mesh to drive a volumetric function, Neural Actor~\cite{neuralActor2021} is conceptually related to our application to faces. In their method, sample points are mapped to a canonical space using the interpolated inverse skinning transform of the closest face of a posed SMPL model. However, an implicit neural deformation function is still learned to account for the low fidelity of the deformation field, as well as for reproducing non-rigid motion. Neural Actor also focuses on whole body deformation only. RigNeRF, a similar method for faces uses an MLP in combination with a 3DMM to improve generalisation~\cite{Athar_2022_CVPR}. However, using an MLP comes with challenges related to run-time and limited generalisation. SnaRF~\cite{chen2021snarf} employs a similar concept without the additional MLP, but instead relies on determining skinning weights for each point in the volume. While a notable work, this method lacks the fidelity required for facial animation.

In terms of using volumetric geometry (e.g. tetrahedral), the use of a guide mesh for smooth deformation fields has been known in geometry editing for a while~\cite{meanValueCoordinatesPaper}. It is relatively uncommon in computer vision and machine learning as compared to learned approaches.

In Neural Cages~\cite{Yifan:NeuralCage:2020}, the authors propose to predict the deformation of a coarse geometric cage and compute the displacement of contained geometry using mean value coordinates~\cite{meanValueCoordinatesPaper}. We see this work as complimentary to our paper in that it would be possible to predict the positions of the vertices of our tetrahedral meshes in canonical pose.

Recently and concurrent to our work, NeRF-Editing~\cite{nerfEditing2022} proposed to use tetrahedral meshes to deform single-frame NeRF reconstructions. While the basic idea of using tetrahedra with barycentric coordinates is the same, our method is more general on several fronts. First, we extend deformation fields to non-tetrahedral primitives such as the mouth region of the face discussed below. Second, we propose an algorithm to make the deformation model real-time using concepts from raytracing. Third, we add a principled way to deal with changes in view direction due to tetrahedral deformation. Fourth, we show that editing a surface mesh and transferring that motion to the volumetric one is not required for good results - we propose several ways of working directly with the tetrahedral geometry such as our volumetric blendshape model, or simulation-based approaches that allows us to demonstrate shattering effects previously elusive to NeRF models.

Prior to this stream of work, the authors of DefTet~\cite{tetDefNvidia} propose deforming a template tetrahedral mesh to fit image observations, where colour and occupancy live directly on the mesh. This was recently extended to learning texture and material directly on mesh surfaces along with environment illumination~\cite{nvdiffrec2021}. Similarly, Deep Marching Tetrahedra proposes to embed a signed distance field (SDF) in a tetrahedral grid, where the SDF is defined directly on mesh vertices~\cite{deepMarchingTetrahedra}.

In the context of medical imaging, Gascón~\etal propose to use tetrahedral deformation for volume data which is similar to this work~\cite{tetDeformationMedical}. However, their focus is mainly on Finite Element Simulation as a means of control and their rasterisation algorithm requires a preprocessing step on the CPU. In contrast, we propose expanded ways to control volumetric models as well as faster tetrahedral lookups.

Because we describe deformations using piece-wise linear tetrahedral primitives, finding out which tetrahedron a point falls in efficiently is key to our method. This problem of localisation in unstructured tetrahedral geometry was previously addressed by~\cite{rtxTetLookupNvidia}, who proposed using hardware-accelerated raytracing in combination with per-face tetrahedral indices in a departure from rasterisation-based techniques~\cite{tetDeformationMedical}. We build on this work by (a) exploiting coherence along rays to reduce the overall number of required rays, and (b) extending the algorithm to work on any closed, non self-intersecting triangulated primitives. 

\label{section:method}
\ready{}
\begin{figure*}[t!]
  \centering
  \includegraphics[width=\textwidth]{{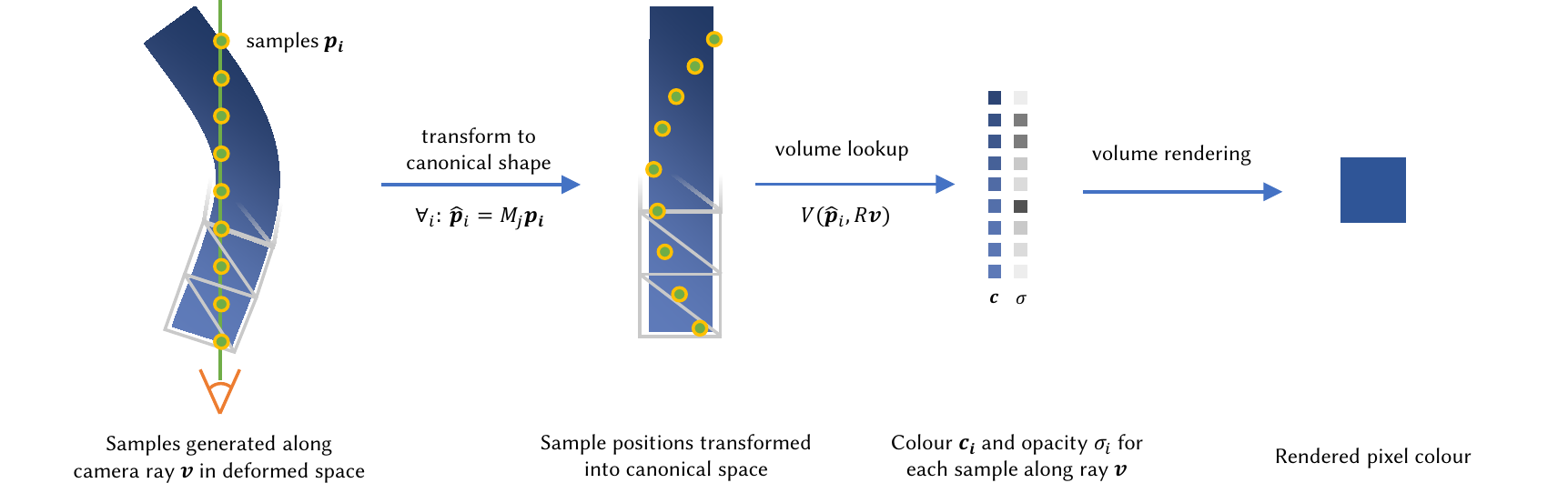}}
  \caption{\ready{} \textbf{Overview of the \methodname{} rendering process.} To render a single pixel a ray is cast from the camera centre, through the pixel into the scene in its deformed state. A number of samples are generated along the ray and then each sample is mapped to the canonical space using the deformation $M_j$ of the corresponding tetrahedron $j$ (Section \ref{sec:tetrahedra}). The volumetric representation of the scene (Section \ref{sec:nerf_basics}) is then queried with the deformed sample position $p'_j$ and the direction of the ray rotated based on the rotation of the $j$-th tetrahedron (Section \ref{sec:view_dir_rotation}). The resulting per-sample density and colour values are then integrated using volume rendering (Equation~\ref{eq:rendering_equation}).}
  \label{fig:pipeline}
\end{figure*}

\section{Method}

\methodname{} allows for controlling volumetric scene representations (Section \ref{sec:nerf_basics}) through deformation of a tetrahedral mesh that envelops the scene in its canonical state. To produce a modified version of the scene one begins by deforming the canonical tetrahedral mesh to the desired shape. The volume rendering method then generates samples along the camera rays in the deformed space and all the samples that fall within the deformed tetrahedral mesh are assigned with the index of the corresponding tetrahedron - this is shown on the left side of Figure  \ref{fig:pipeline}, details in Section \ref{sec:point_lookups}. Thanks to the known sample-tetrahedron associations, the samples can be transformed to the canonical space where the scene representation lives (Figure \ref{fig:pipeline} and Sections \ref{sec:tetrahedra}, \ref{sec:view_dir_rotation}). Once the canonical state of each sample is known volume rendering can be applied as normal.

\methodname{} is agnostic to the way in which the tetrahedral deformation is generated and what kind of an object is deformed. In Section \ref{sec:ext_generic_objects} we describe how our method can deform generic objects while in Section \ref{sec:animatable_faces} we describe its application to faces. Finally, \methodname{} is also flexible in terms of the type of training sequence as it can work both with sets of images of a static scene as well as sequences where each image represents the scene in a different state, details in Section \ref{sec:enrolment}.


\subsection{Learned Volumetric Scene Representations}
\label{sec:nerf_basics}
\ready{}
Volumetric scene representations using learned functions to account for the scattering events, 
have enjoyed a renaissance in recent years 
The final colour, opacity and depth at each pixel is obtained by sampling points $p_i$ along rays traced through a bounded 3D scene. For each sample a function $\mathcal{V}$ is queried to obtain the colour $\bm{c}$ as well as density $\sigma$ at that position in space. Commonly, the colour of a pixel on the image plane, $\bm{\hat{c}}$, is obtained via volume rendering using an emission-absorption form of the volume rendering equation
\cite{volumeRenderingEquation}: 
\begin{equation}
    \bm{\hat{c}} = \sum^{N}_{i=1} w_i \bm{c}_i, \quad w_i = T_i (1-\exp (-\sigma_i \delta_i)), \label{eq:rendering_equation}
\end{equation}
where $\delta_i= \left( \bm{p}_{i+1} - \bm{p}_i \right)$ denotes the distance between samples (in total $N$) along straight rays, and the transmittance, $T_i$, is defined as $T_i = \exp (-\sum_{j=i}^{i-1} \sigma_j \delta_j)$. 
$\mathcal{V}$ is usually modelled by a Multi-Layer Perceptron (MLP) \cite{nerf, instantNGP}, an explicit voxel grid \cite{plenoctrees, directVoxelOptimisation} or a combination of both \cite{neuralSparseVoxelFields, SnergBaked}.
In addition to sample position $\bm{p}$, $\mathcal{V}$ is also conditioned on the direction of the ray $\bm{v}$, which allows it to model view-dependent effects such as specular reflections.
Models such as NeRF also typically omit image formation models for traced rays, opting instead to evaluate one ray per pixel, located at evenly spaced intervals. The resulting aliasing can be addressed by integrating the frequency encodings \cite{mipNeRF} as compared to more expensive oversampling.

\subsection{Modelling Volumetric Deformations with Closed Triangular Primitives}
\ready{}
\label{sec:tetrahedra}
\begin{figure}
  \centering
  \includegraphics[width=0.9\columnwidth]{{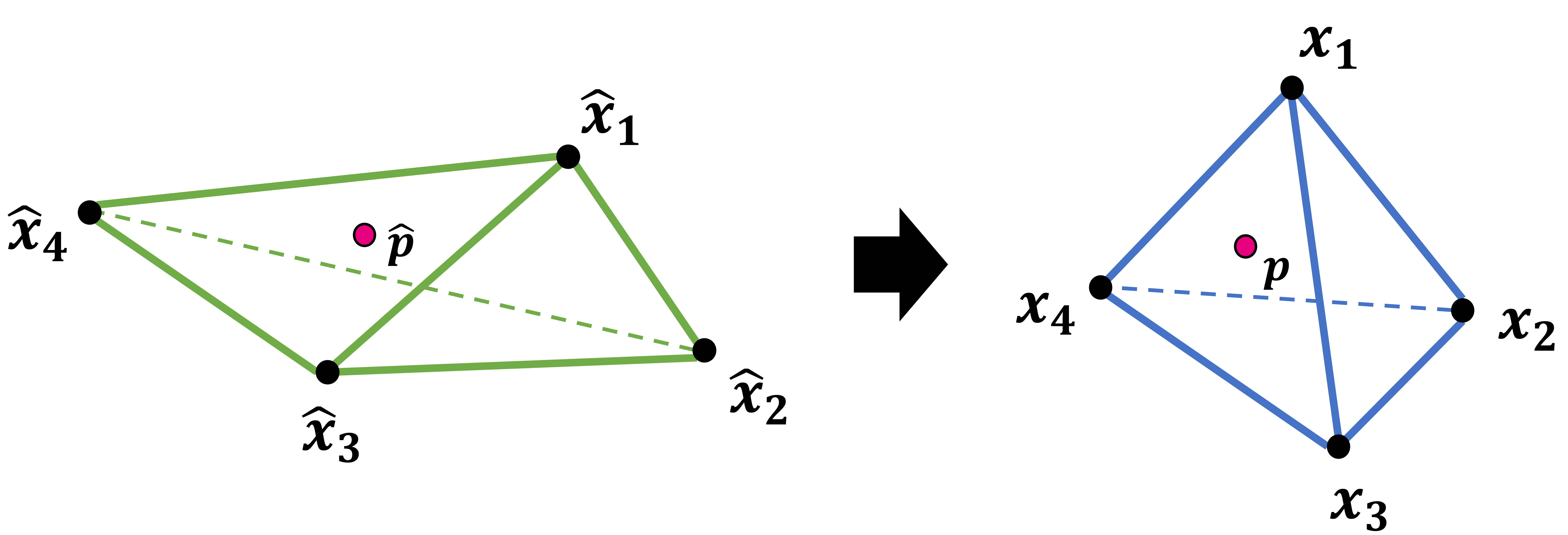}}
  \caption{\ready{} A point $\bm{p}$ in deformed spaced is mapped to $\bm{\hat{p}}$ in canonical space using barycentric coordinates defined for both the canonical tetrahedron $X = \{\bm{x}_1, \bm{x}_2, \bm{x}_3, \bm{x}_4\}$ as well as the deformed tetrahedron $\hat{X} = \{\hat{\bm{x}}_1, \hat{\bm{x}}_2, \hat{\bm{x}}_3, \hat{\bm{x}}_4\}$.}
  \label{fig:TetrahedralDeformationConcept}
\end{figure}
Following prior work, our approach to animation relies on mapping points to positions in a `rest' or `canonical' space. As NeRF-based models use volumetric representations, this requires mapping points in the volume that supports the radiance field. In order to animate a wide variety of object categories in this representation interactively, the deformation model should (a) be real-time, (b) capable of representing both smooth and discontinuous functions, (c) allow for intuitive (manual) control, thereby allowing good extrapolation or generalisation to configurations not observed in training.

We propose to represent complex motion fields using closed triangular volumetric primitives. The most common type of primitive used in this work is a tetrahedron, and we start our explanation from this perspective.
This representation is compatible with GPU-accelerated ray tracing and can be queried in milliseconds, even with complex geometry. It is capable of reproducing hard object boundaries by construction and can be edited in off-the-shelf software thanks to being composed of only points and triangles.

A tetrahedron, one fundamental building block of our method, is a four-sided pyramid. We define the undeformed `rest' position of its four constituent points as:
\begin{equation}
    X = \{\bm{x}_1, \bm{x}_2, \bm{x}_3, \bm{x}_4\},
\end{equation}
and denote the deformed state $\hat{X} = \{\hat{\bm{x}}_1, \hat{\bm{x}}_2, \hat{\bm{x}}_3, \hat{\bm{x}}_4\}$. Because tetrahedra are simplices, we can represent points that fall inside them using barycentric coordinates ($\lambda_1, \lambda_2, \lambda_3, \lambda_4$) with respect to $X$ or $\hat{X}$~\cite{femSimReference}.

An input point $\hat{\bm{p}}$ can be parameterised as $ \hat{\bm{p}} = \sum_{i=1}^{4} \lambda_i * \hat{\bm{x}}_i$ if it falls inside the tetrahedron, and we can obtain its rest position $\bm{p}$ in the canonical space as:
\begin{equation}
    \bm{p} = \sum_{i=1}^{4} \lambda_i * \bm{x}_i.
\label{eq:barycentricPointLookup}
\end{equation}
This is illustrated in Figure~\ref{fig:TetrahedralDeformationConcept}.
For volumetric primitives that are not simplices (such as the interior of the mouth of our face model or any irregularly shaped rigid part of a mesh), no such barycentric mapping can be defined. 
We use simple affine transformations in these cases, which are expressive enough for large rigidly moving sections of the motion field.

\subsection{Accounting for Changes in View Direction}
\label{sec:view_dir_rotation}
\ready{}
Transforming sample positions between canonical and deformed spaces takes care of changes to the shape of the scene. However, as described in Section~\ref{sec:nerf_basics} the density and colour at each point in the scene is a function of both sample position and view direction. If sample positions are moved, but view directions stay unchanged, the light reflected off the elements of the scene will appear the same for every deformation.
To alleviate this problem we propose to rotate the view direction $\bm{v}$ of each sample with a rotation $R$ between the canonical tetrahedron and its deformed equivalent:

\begin{align}
\begin{split}
\bm{\hat{v}} &= R \bm{v}, \\
U, E, V &= SVD((X - \bm{c}_x)^T(\hat{X} - \bm{c}_{\hat{x}})), \\
R &= UV^T,
\end{split}
\end{align}
where $\bm{c}_x, \bm{c}_{\hat{x}}$ are the centroids of the canonical and deformed states of the tetrahedron that a given sample falls into. With this approach, the direction from which the light is reflected at each point of the scene will match the deformation induced by the tetrahedral mesh. Note however, that the reflected light will represent the scene in its canonical pose.

In practice, computing $R$ for each sample or even each tetrahedron in the scene is inefficient as it requires computing Singular Value Decomposition (SVD) for each $R$ separately. What we do instead is take a stochastic approach where we compute $R$ for a small fraction $\rho$ of all tetrahedra and propagate $R$ to the remaining tetrahedra via nearest neighbour interpolation. In all experiments we set $\rho=0.05$.

\subsection{Enrolling from as Little as a Single Frame}
\label{sec:enrolment}
\ready{}
As we decouple the deformation of the model from its appearance in canonical space, we can train our dynamic scene representations on images of a scene taken from multiple viewpoints at the same moment in time. This is not a requirement as we can also enroll from images taken at different moments in time as long as we have the ground-truth deformation for these frames. Ground truth deformations are usually only available for synthetic scenes, for real data we need to resort to object trackers to estimate the deformation and account for tracker failures. Small errors in the tracker produce small errors in the deformation into the canonical space, where the scene's appearance is learned, and this in turn produces blurrier renders because the canonical space averages the appearance of samples within regions defined by these errors. While learning the errors in all training frames is a possibility, we would still need to account for temporal changes to appearance caused by illumination and other factors.
We have empirically found that training from a single frame where a good estimate of the deformation is available produces sharp images and simplifies the method. 

\subsection{Tetrahedral Point Lookups}
\label{sec:point_lookups}
\ready{}
While the proposed deformation model is simple, it relies on accurately determining which primitive a point falls into.

With complex meshes, checking each tetrahedron for association with each input point we wish to deform naively is not feasible given the complexity of point-in-tetrahedron tests.
We choose triangles as primitives of our motion field to build on the work of \cite{rtxTetLookupNvidia}, who show that for non self-intersecting tetrahedral meshes the notions of a point being `in front' or `behind' a certain triangle are uniquely determined by the triangle vertices' winding order.
In their approach determining which tetrahedron a point belongs to amounts to shooting a ray in a random direction from the point, evaluating the triangle at first intersection and checking which side of the triangle the sample is on. This identifies the tetrahedron uniquely as each triangle can belong to at most two tetrahedra. Especially when hardware acceleration is available, these queries are highly efficient in terms of memory and compute.

We extend this work in two ways: First, we apply the same acceleration to arbitrarily triangulated shapes that let us combine tetrahedra with triangulated rigidly-moving shapes that do not need to be filled with tetrahedra but can be treated as a unit in terms of deformation.  Second, we reduce the number of point-in-tetrahedron tests required by noting that many samples along a single ray can fall into the same element. If we know the previous and next intersection, a simple depth-test determines which tetrahedron samples fall into. Barycentric coordinates are linear, and so we can obtain a barycentrically interpolated value by interpolating values at the previous and next intersection within each element. To do this, we rewrite Equation~\eqref{eq:barycentricPointLookup} as:
\begin{equation}
    \bm{p} = \alpha * \sum_{i=1}^{4} \lambda_i^1 * \bm{x}_i^1 + (1.0 - \alpha) * \sum_{i=1}^{4} \lambda_i^2 * \bm{x}_i^2,
\end{equation}
where the superscripts $1$ and $2$ refer to the previous and next intersection, and $\alpha$ is the normalised distance between the two intersections which defines the point that we are interpolating for.
Thanks to this modification, per \textit{point} values remain stable even if the `wrong' side of a triangle (or incorrect triangle all together) is queried due to a lack of numerical precision. This is because intersections tend to be reported as numerically close as possible to \textit{triangles} in ray tracing libraries such as Optix \cite{optixMainPaper}. One important side effect of this per-ray as opposed to per-point formulation of tetrahedral index lookups is that it naturally integrates with ray marching approaches to rendering, such as FastNeRF \cite{garbin2021fastnerf}. In the latter, rays are terminated based on transmittance, which our reformulated tetrahedral lookup algorithm naturally allows.

\section{Applications}

\subsection{Applying \methodname{} to support generic objects}
\label{sec:ext_generic_objects}
\ready{}
\begin{figure*}[h]
  \centering
  \includegraphics[width=\textwidth]{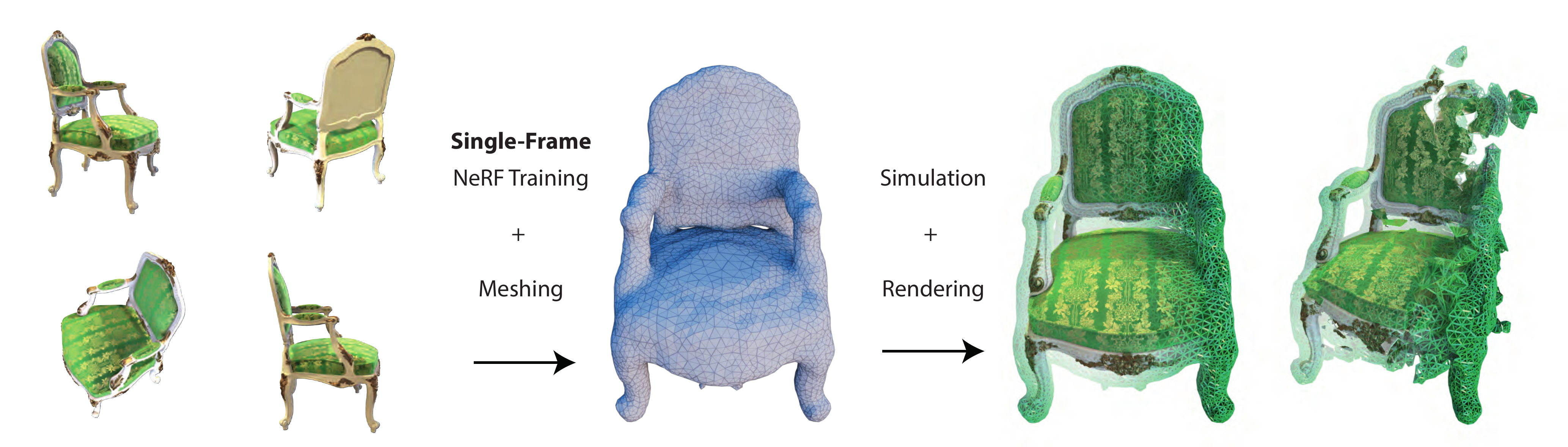}
  \caption{\ready{} Converting a static scene to an animatable one. Surface mesh extraction (from volume density), tetrahedralisation and simulation are automated using off-the-shelf software, but manual animation is equally possible. }
  \label{fig:animatingGenericScenes}
\end{figure*}
Given a collection of images and associated camera intrinsic and extrinsic parameters, we can use \methodname{} to animate generic static scenes. The volumetric model is trained using the standard approach of the given scene representation method (we use FastNeRF~\cite{garbin2021fastnerf} as detailed in Section~\ref{section:implementation}).
To create the volumetric geometry, we first extract a coarse mesh from the density of a trained model using Marching Cubes~\cite{marchingCubes}, and subsequently produce a tetrahedral embedding of that mesh using off-the-shelf software (SideFX Houdini). Once we have a coarse mesh we can animate it or produce interesting motion with simulation, letting the object deform or shatter under the effects of gravity. This process is shown in Figure~\ref{fig:animatingGenericScenes}, and is also done using off-the-shelf software (SideFX Houdini\footnote{https://www.sidefx.com/products/houdini/}). We note that the function deforming the tetrahedra or assembling the tetrahedra itself could be learned as well~\cite{tetDefNvidia, Yifan:NeuralCage:2020}, but we leave this as future work.

\subsection{Applying \methodname{} to support Faces}
\label{sec:animatable_faces}

\ready{}
To achieve generalisation and control of faces with \methodname{}, we introduce a generalisation of parametric 3DMM, which animates a mesh with a skeleton and blendshapes, to a parametric model that animates a volume around a mesh which we coin Vol3DMM. Vol3DMM animates a volumetric mesh using a set of volumetric blendshapes and a skeleton.

\begin{figure*}[h]
  \centering
  \includegraphics[width=0.85\textwidth]{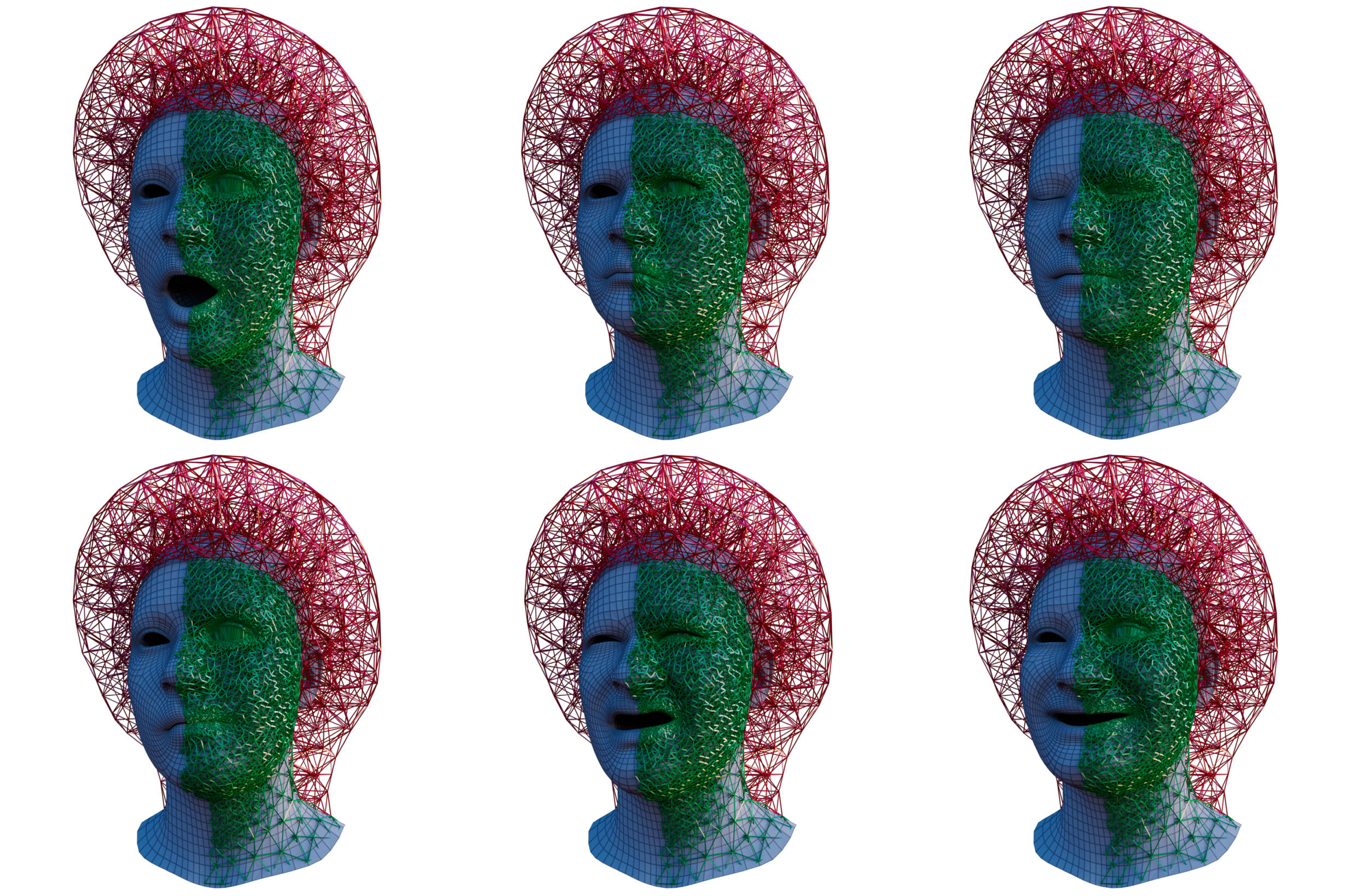}
  \caption{\ready{} Our volumetric face model extends the blendshapes of a traditional face 3DMM model from a surface (blue mesh) to a tetrahedral partition of the volume around it (green). The tetrahedral volume defines the support of the deformation and can be extended to cover hair, headphones or headgear (red).}
  \label{fig:FaceVolumetricBlendshapes}
\end{figure*}

We define the skeleton and blendshapes of Vol3DMM by extending the skeleton and blendshapes of a parametric 3DMM face model of~\cite{fakeItTillYouMakeIt}. The skeleton has the same four bones as the 3DMM face model: the root bone controlling global rotation, the neck, the left eye, and the right eye. To use this skeleton in Vol3DMM, we extend the linear blend skinning weights from the vertices of the 3DMM mesh to the vertices of the tetrahedra by a simple nearest-vertex look up, that is, each tetrahedron vertex has the skinning weights of the closest vertex in the 3DMM mesh. The volumetric blendshapes are also created by extending the 224 expression blendshapes and the 256 identity blendshapes of the 3DMM model to the volume surrounding its template mesh: the $i$-th volumetric blendshape of Vol3DMM is created as a tetrahedral embedding of the mesh of the $i$-th 3DMM blendshape. To create the tetrahedral embedding, we use the same procedure used to create a single volumetric structure from a generic mesh described in Section~\ref{sec:ext_generic_objects} and manually curate it to avoids tetrahedral inter-penetrations between upper and lower lips, covers hair, and have higher resolution in areas subject to more deformation. The resulting Vol3DMM is more easily understood graphically, as illustrated in Figure~\ref{fig:FaceVolumetricBlendshapes}.

As a result of this construction, Vol3DMM is controlled and posed with the identity, expression, and pose parameters $\alpha, \beta, \theta$ of the 3DMM face model. This means that we can animate it with a face tracker built on the 3DMM face model by changing $\beta, \theta$ and, more importantly, that it generalises convincingly to any expression representable by the 3DMM face model. During training we use the parameters $\alpha, \beta, \theta$ to pose the tetrahedral mesh of Vol3DMM to define the physical space, while the canonical space is defined for each subject by posing Vol3DMM with identity parameter $\alpha$ and setting $\beta, \theta$ to zero for a neutral pose. Each tetrahedron in the physical space $\lbrace \hat{X}^i \rbrace_{i=1}^{n}$ has a canonical counterpart $\lbrace X^i \rbrace_{i=1}^{n}$. From each camera, we shoot rays in the physical space, detect the tetrahedron $\hat{X}^k$ incident to each sample $\hat{\bm{p}}$ along the ray and compute its barycentric coordinates $(\lambda^k_1, \lambda^k_2, \lambda^k_3, \lambda^k_4)$ such that
\begin{align}
\hat{\bm{p}}=\sum_{i=1}^{4} \lambda^k_i \hat{\bm{x}}^k_i,
\end{align}
where $(\hat{\bm{x}}^k_1, \hat{\bm{x}}^k_2, \hat{\bm{x}}^k_3, \hat{\bm{x}}^k_4)$ are the vertices of the tetrahedron $\hat{X}^k$. The samples inside the volume are then deformed into canonical space as $\bm{p}=\sum_{i=1}^{4} \lambda^k_i \bm{x}^k_i$ and used as input to the coordinate-based MLP that approximates the neural radiance field in canonical space. Once trained, the avatar for this subject is controlled by pose and expression parameters of the 3DMM model, which can be obtained by a real-time face tracking system~\eg~\cite{wft}. In particular, our 224 expression blendshapes are a superset of the 85 expression blendshapes our real-time face tracker. This allows us to build a volumetric model compatible with a real-time face tracker but not limit its expressivity to a specific set of blendshapes chosen for a real-time system.

\textbf{Support for mouth interior}.
\begin{figure}[h]
  \centering
  \includegraphics[width=0.35\textwidth]{{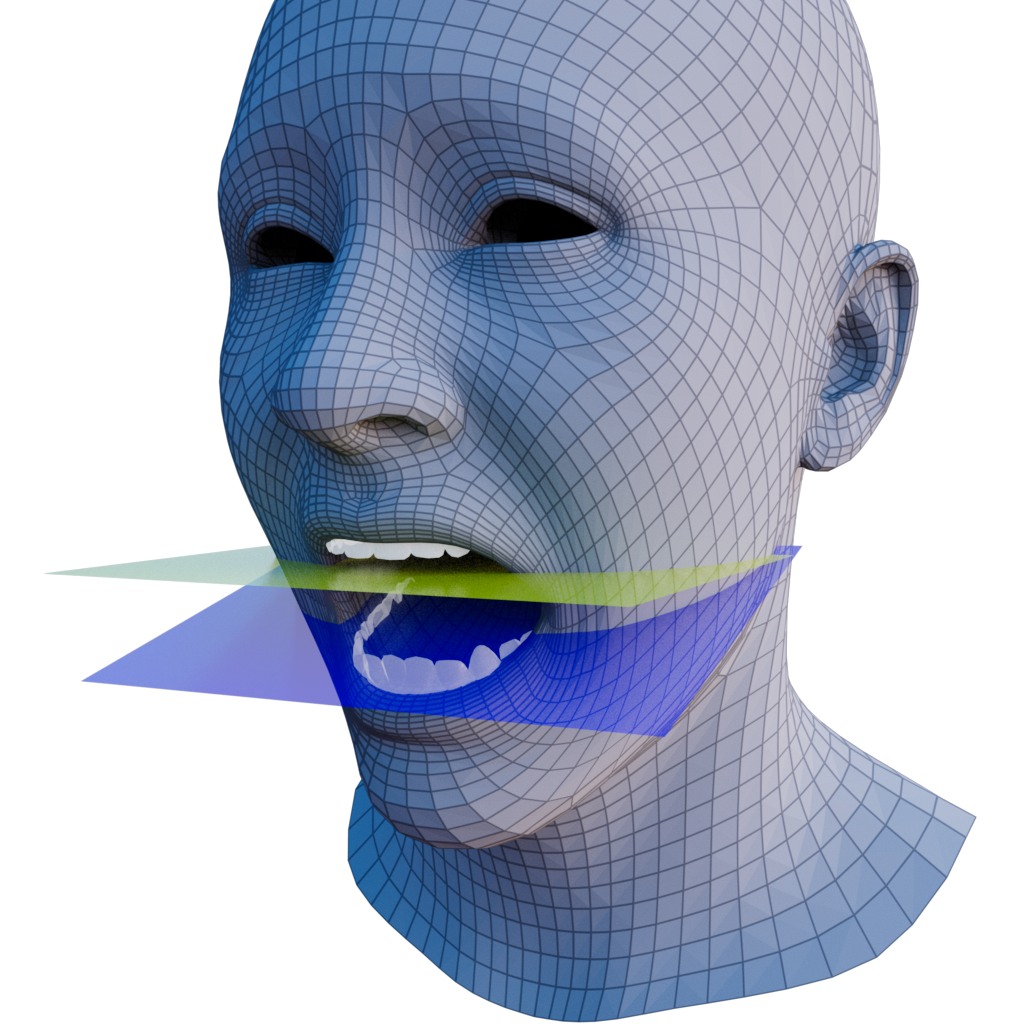}}
  \caption{The two planes delineating rigid mouth regions. Tetrahedral mesh not shown for clarity, reference teeth geometry for illustration only.}
  \label{fig:mouth_planes}
\end{figure}
Vol3DMM is built on top of a 3DMM - a surface model of the face that does not handle the inside of the mouth.
Simply filling the mouth with tetrahedra and extrapolating the deformation of the surface model to those elements would result in unrealistic motion, especially for rigid components such as the teeth. Since we animate a static volumetric model, we cannot learn a deformation model from temporal data. Instead, we model the mouth interior as a closed, irregularly shaped triangular mesh. Samples in its interior are deformed using two rigid `jaw' regions delineated by planes, one placed just below the top teeth and one just above the bottom teeth (see Figure~\ref{fig:mouth_planes}). Samples above the top plane are moved by the motion of the head, while samples below the bottom plane are moved together with the jaw.
We decide not to support the tongue at this time, and therefore assume the space between the planes is empty and we do not render the samples that fall in that region.

\section{Implementation}
\label{section:implementation}

We use a solution similar to FastNeRF~\cite{garbin2021fastnerf} as its high performance allows the whole system to run in real-time on a Nvidia RTX 3090 GPU. Compared to FastNeRF, we increase the layer width of the position-dependent network to 512 and the number of frequency encodings of that network to 12 with the goal of improving render quality. We also reduce the size of the view-dependent network to 2 layers consisting of 8 units each to mitigate overfitting.
Finally we modify FastNeRF's rendering infrastructure by substituting the density-derived collision mesh with the tetrahedral geometry of Section~\ref{sec:tetrahedra} with 8908 vertices and 39179 tetrahedra for the faces and lower resolution meshes for the various synthetic scenes. 

We use the Adam optimizer~\cite{kingma2014adam} to train our models with a least-squares RGB loss. For faces we add a sparsity-promoting term to the loss. 
\ready{}

\textbf{Sparsity terms for faces}. We use additional sparsity terms to deal with incorrect background reconstruction and reduce artefacts arising from disocclusions\footnote{Disocclusion is a situation where a previously occluded object becomes visible.} in the mouth interior region. We use the following Cauchy loss on the density $\sigma$ along rays, as in SNeRG~\cite{SnergBaked}: 

\begin{equation}
    \mathcal{L}_{s} = \frac{\lambda_{s}}{N} \sum_{i, k} \log \left( 1 + 2\sigma \left( \textbf{r}_{i} \left( t_{k} \right) \right)^{2} \right),
\end{equation}

where $i$ indexes rays $\textbf{r}_{i}$ shot from the training cameras, $k$ indexes samples $t_{k}$ along each of the rays, $N$ is the total number of samples, and $\lambda_{s}$ is a scalar hyperparameter. To ensure the space is evenly covered by the sparsity loss, we only apply it to the coarse samples. In our experiments we use 128 coarse samples, 64 fine samples.

We apply the sparsity loss in two regions: in the volume surrounding the head and in the mouth interior. Applied to the volume surrounding the head, the sparsity loss prevents opaque regions appearing in areas where there is not enough multi-view information to disentangle foreground from background in 3D. To detect these regions, we apply the loss to (1) samples which fall in the tetrahedral primitives as this is the only region we render at test-time, and (2) samples which belong to rays which fall in the background in the training images as detected by 2D face segmentation similar to~\cite{fakeItTillYouMakeIt} of the training images.

We also apply the sparsity loss to the coarse samples that fall inside the mouth interior volume. This prevents the creation of opaque regions inside the mouth cavity in areas that are not seen at training, and therefore have no supervision, but become disoccluded at test time.

We set $\lambda_{s}=10^{-4}$ for the volume surrounding the head and use $\lambda_{s}=2 \times 10^{-6}$ for the mouth interior.

\ready{}\textbf{Disoccluded mouth interior colour}. The sparsity loss inside the mouth interior ensures there is no unnecessary density inside the mouth interior. However, the colour behind the regions which were occluded at training time remains undefined, resulting in visual artefacts when these regions are disoccluded at test-time. We mitigate this by overriding the colour and density of the last sample along each ray that falls in the mouth interior. We set the rendered colour to match the colour visible between the teeth in the training frame.

\begin{figure}
  \centering
  \includegraphics[width=0.9\columnwidth]{{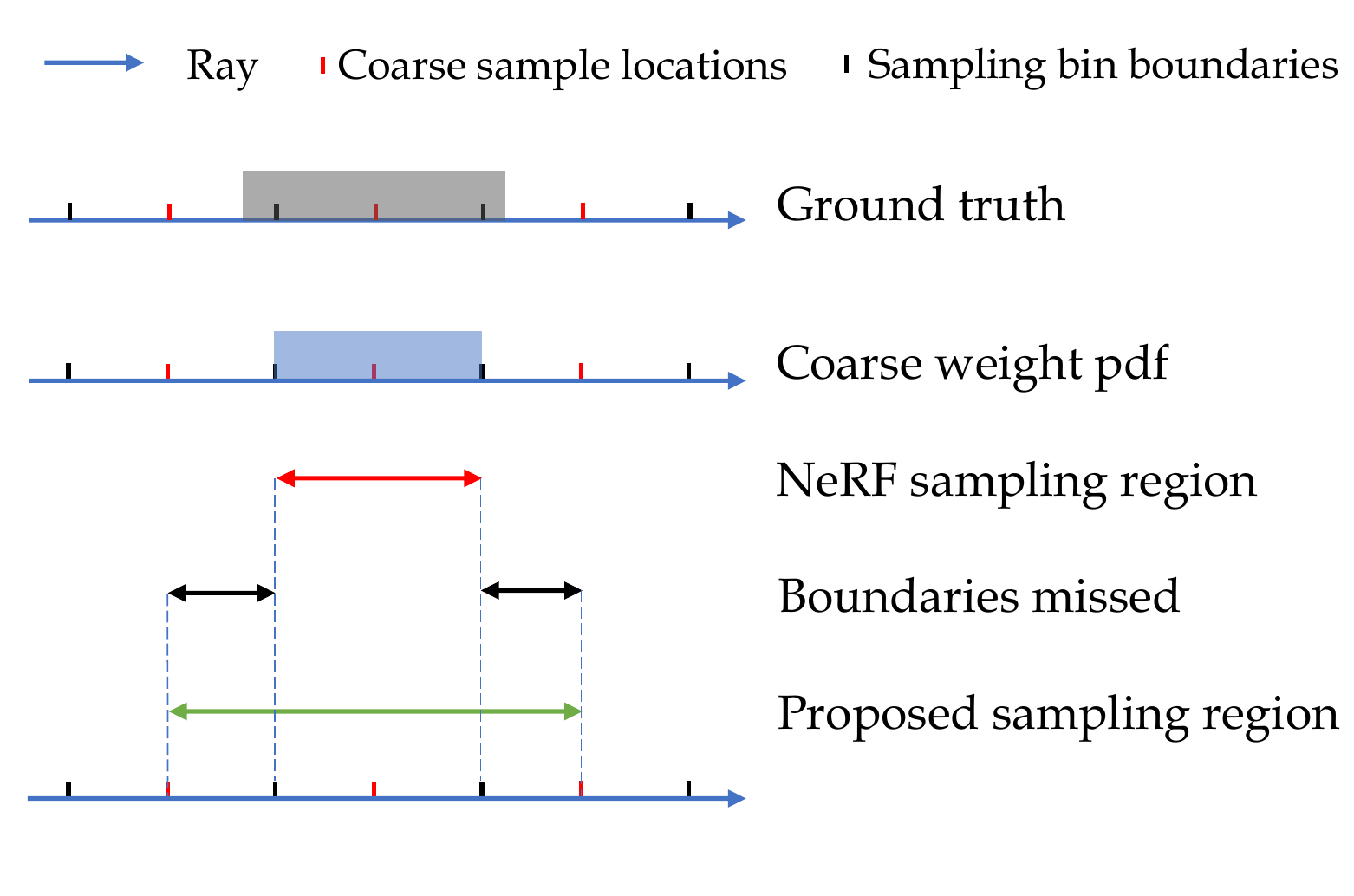}}
  \caption{NerF's per-ray sampling aims to accurately capture the ground truth density (top). The first set of samples (red ticks) are used to obtain a coarse estimate. The second set of samples is placed according to the estimated contribution to the ray colour (second row, blue). NeRF's second round of sampling (third row, red arrow) takes samples from bins defined by mid-points between the coarse samples (black ticks). This leads to errors (third row, black arrows) when using the same network for the coarse and fine samples. We propose to extend the boundaries of the sampling bins (third row, green arrow) to avoid these artefacts.}
  \label{fig:03_method/sampling/IS_method_fig}
\end{figure}

\subsection{Sampling}
\label{sec:sampling}
NeRF~\cite{nerf} uses both a coarse and a fine MLP to approximate opacity functions at coarse and fine resolutions. First, $N_c$ samples are evaluated by the first network to obtain a coarse estimate of the densities along the ray. This estimate guides a second round of $N_f$ samples, placed according to the normalised weights $\hat{w_i} = w_i / \sum_i w_i$ (with $w_i$ defined as in Equation~\ref{eq:rendering_equation}) of the coarse stage. The fine network is then queried at both coarse and fine sample locations, leading to $N_c$ evaluations in the coarse network and $N_c+N_f$ evaluations in the fine network. During training, both MLPs are optimised independently, but only the samples from the fine one contribute to the final pixel colour.

We would like to to improve efficiency by avoiding querying the fine network at the locations of coarse samples and instead reusing the output from the first round of coarse samples. Using two networks, this would result in the coarse network modelling the overall appearance of the scene and the fine network only the areas of high density. Since we use FastNeRF, a single cache has to be distilled for the whole scene. With two MLPs, there isn't a single network that encodes both  high-frequency detail and covers all required space. Thus, we use a single network for both sampling stages.

We found that making the change to a single network without other adjustments leads to artefacts due to the way the reference implementation of NeRF's hierarchical sampling works. As shown in Figure~\ref{fig:03_method/sampling/IS_method_fig}, areas around segments of a ray that have been assigned high weights can be clipped. Clipping can occur because the bin placement for drawing the fine samples treats density like a step function at the sample location instead of a point estimate of a smooth function.
We hypothesise \cite{nerf} did not observe this because the coarse  network learns a lower-frequency function due to never being sampled at the refined locations. Our solution is to increase the size of bins used in the fine sampling stage. This makes an assumption of the density being smooth, which is in line with recent findings on volumes with exponential transmittance \cite{nonExponentialScattering}.

We note that addressing sample efficiency and aliasing in Neural Radiance Fields has been the topic of other works, e.g. Mip-NeRF~\cite{mipNeRF}. 
Our changes are complementary to this approach and should be applicable to most NeRF-like methods using two-stage sampling.

\section{Experiments}
\label{section:experiments}
\ready{}
We conduct evaluation on both synthetic data with easy to interpret, procedural motion to motivate the need for our method, as well as more challenging real data of human faces using blendshape animation. Note that human faces are a particularly difficult case due to a non-trivial combination of rigid and (visco)elastic motion. Ablations further help explain the motivation for design choices. 

Please note that all timings were evaluated on a single RTX 3090 GPU, and that our models are implemented in pytorch with custom extensions for tetrahedral lookups and FastNeRF, following the implementation of \cite{garbin2021fastnerf}.

\subsection{Synthetic Data}

\subsubsection{Baseline Comparison}
\label{sec:synth_baseline_comp}
To demonstrate why explicit volume deformation methods such as ours are needed despite the existence of seemingly powerful coordinate-based deformation models, we first create a simple dataset of a propeller undergoing a continuous compression and rotation. The compression motion can be thought as a very simple linear blendshape model, whereas rotation is part of joint-based animation models. For both types of deformation, we render $48$ consecutive temporal frames for $100$ cameras. These frames do not exhibit random motion. Rather, successive frames can be thought of as successive time steps of a realistic animation representing small increments per frame of the respective type of motion.

\begin{figure}
  \centering
  \includegraphics[width=0.23\textwidth]{{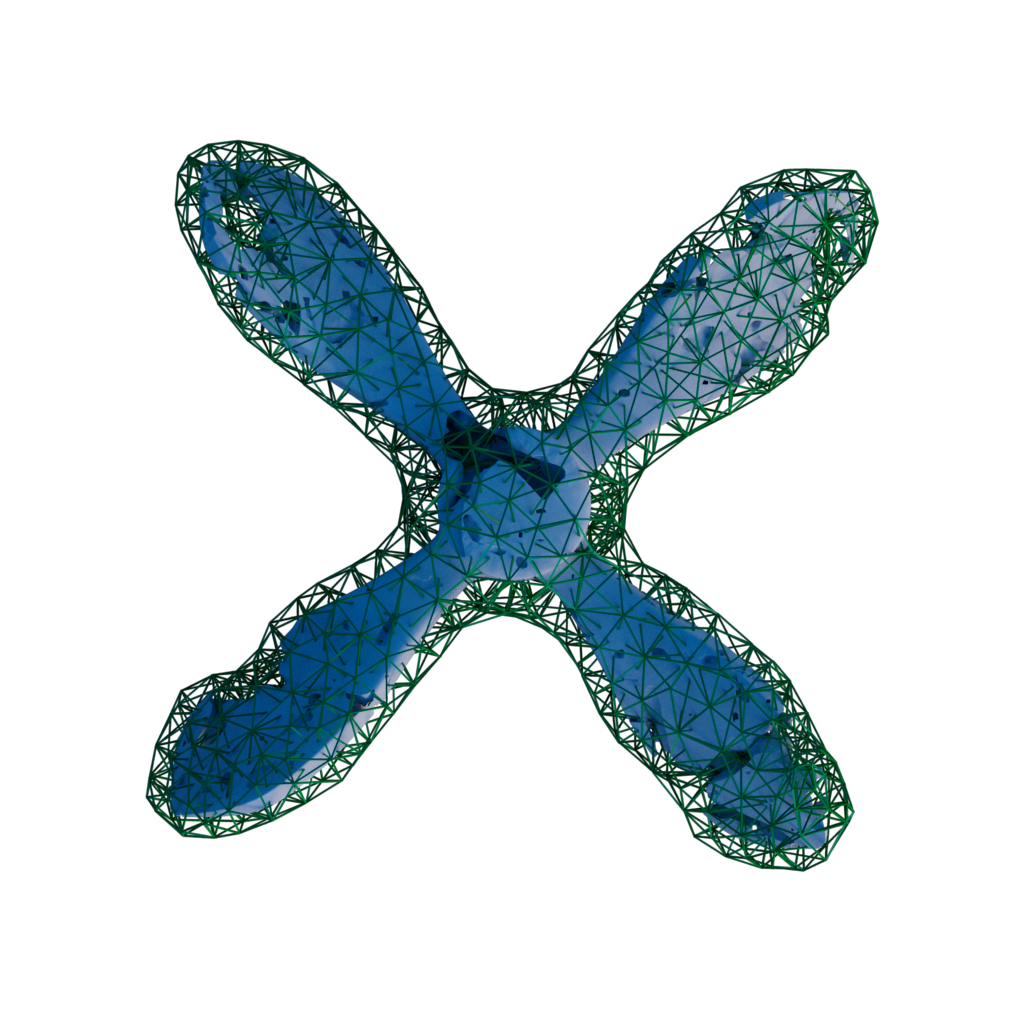}}
  \includegraphics[width=0.23\textwidth]{{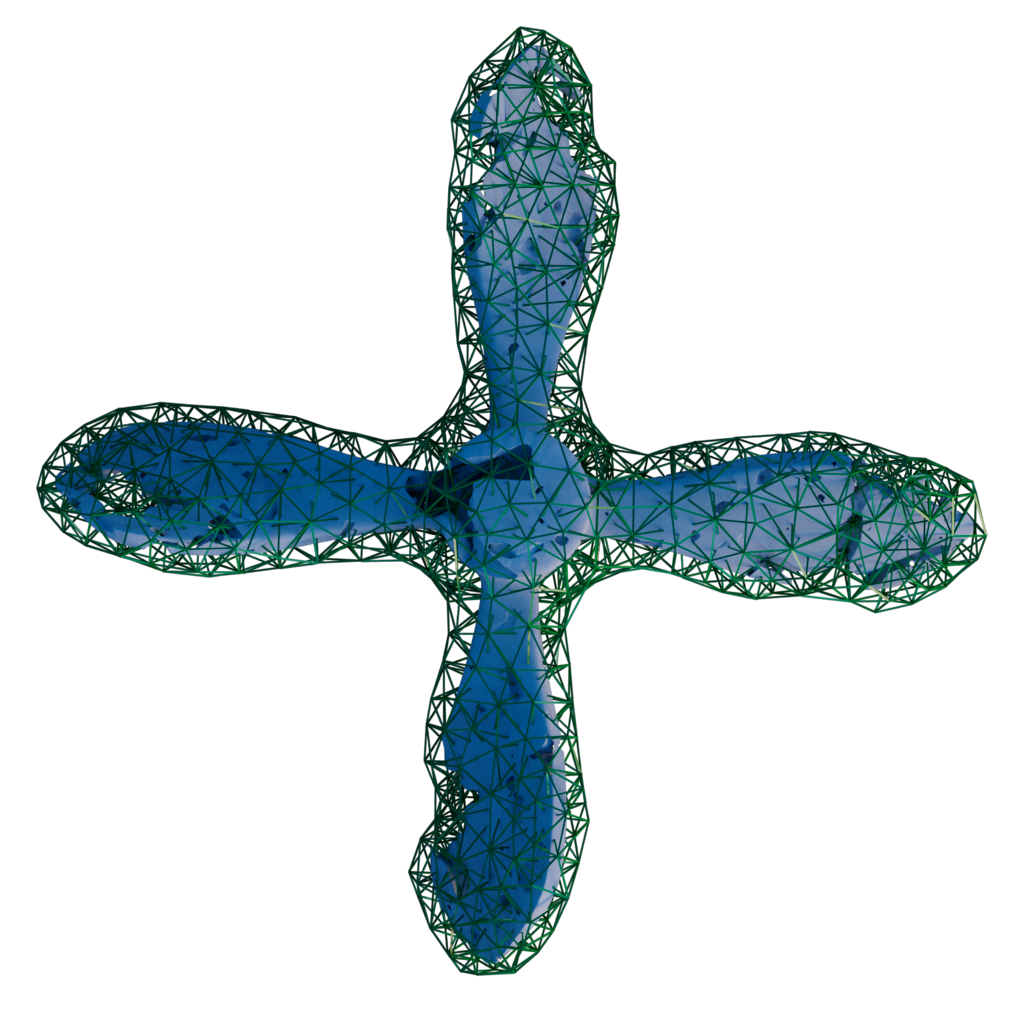}}
  \caption{\ready{} Coarse Tetrahedral Cage (roughly 2K elements) of the 'propeller' dataset, in 'rest` state of the left, and rotated counter clockwise by $45$ degrees on the right. Propeller mesh shaded for illustration only.}
  \label{fig:propeller_tet_cage}
\end{figure}

\begin{figure}[t]
    \makebox[2.2cm]{\centering D-NerF } 
    \makebox[2.2cm]{\centering Ours}
    \makebox[2.2cm]{\centering GT}\\
    \begin{subfigure}[c]{1.0\linewidth}
    \centering\includegraphics[width=0.8\columnwidth]{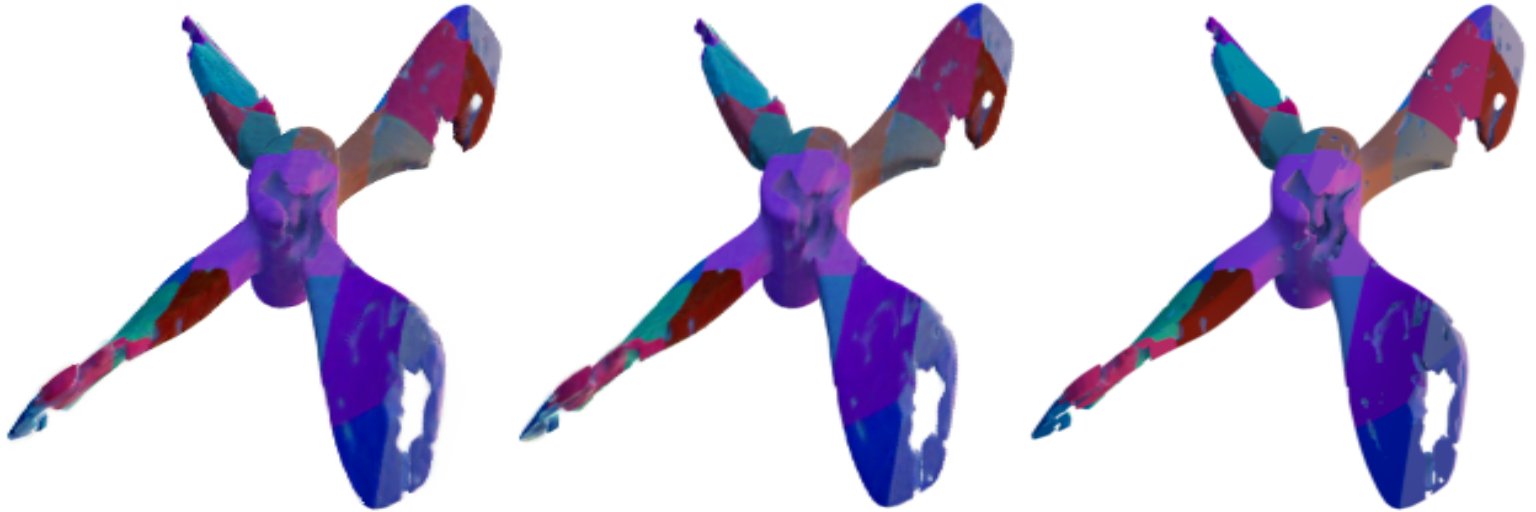}
    \caption{Interpolation: horizontal compression}
    \end{subfigure} \\
    \begin{subfigure}[c]{1.0\linewidth}
    \centering\includegraphics[width=0.8\columnwidth]{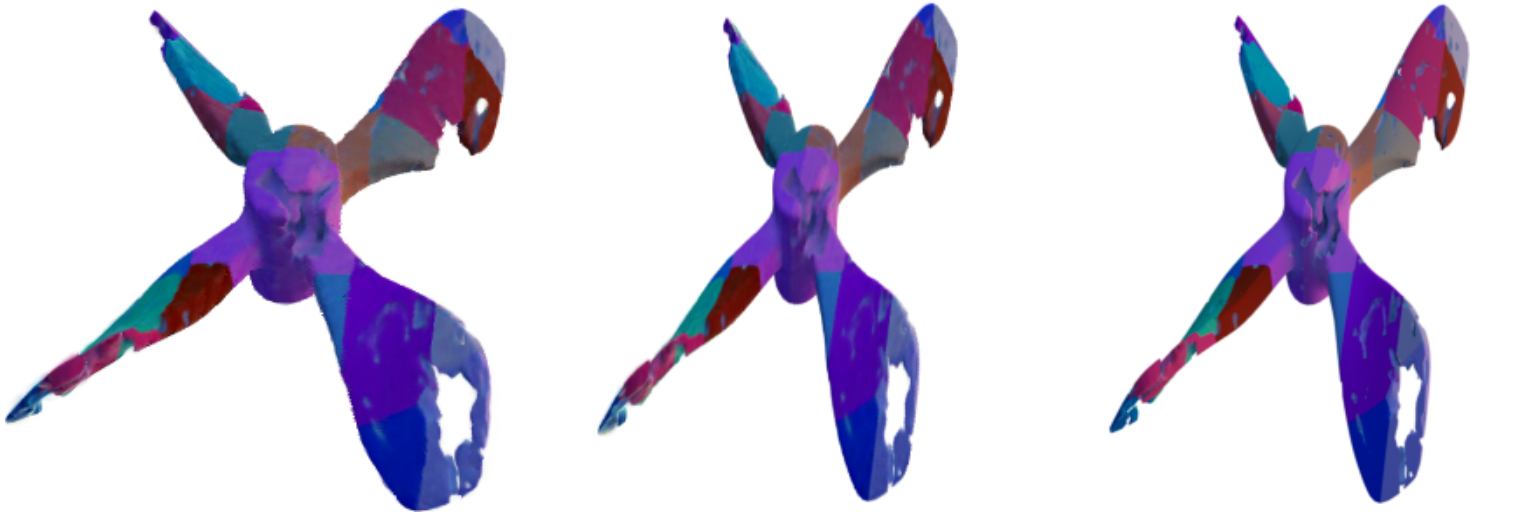}
    \caption{Extrapolation: horizontal compression}
    \end{subfigure} \\
    \begin{subfigure}[c]{1.0\linewidth}
    \centering\includegraphics[width=0.9\columnwidth]{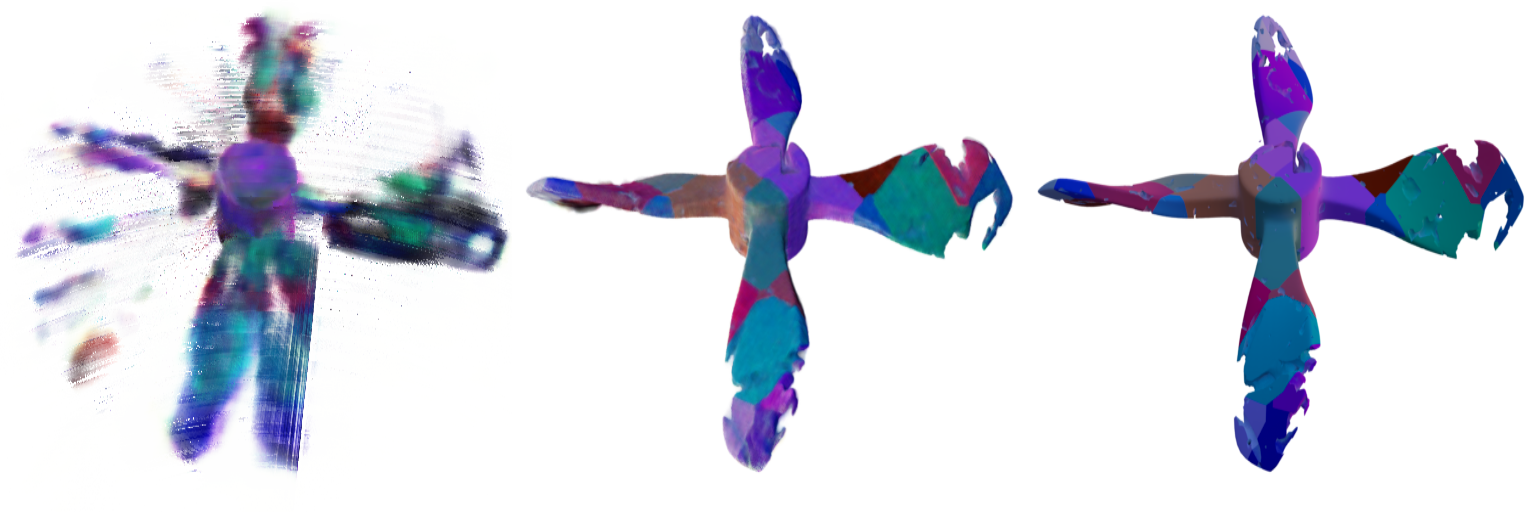}
    \caption{Interpolation: rotation}
    \end{subfigure} \\
    \begin{subfigure}[c]{1.0\linewidth}
    \centering\includegraphics[width=1.0\columnwidth]{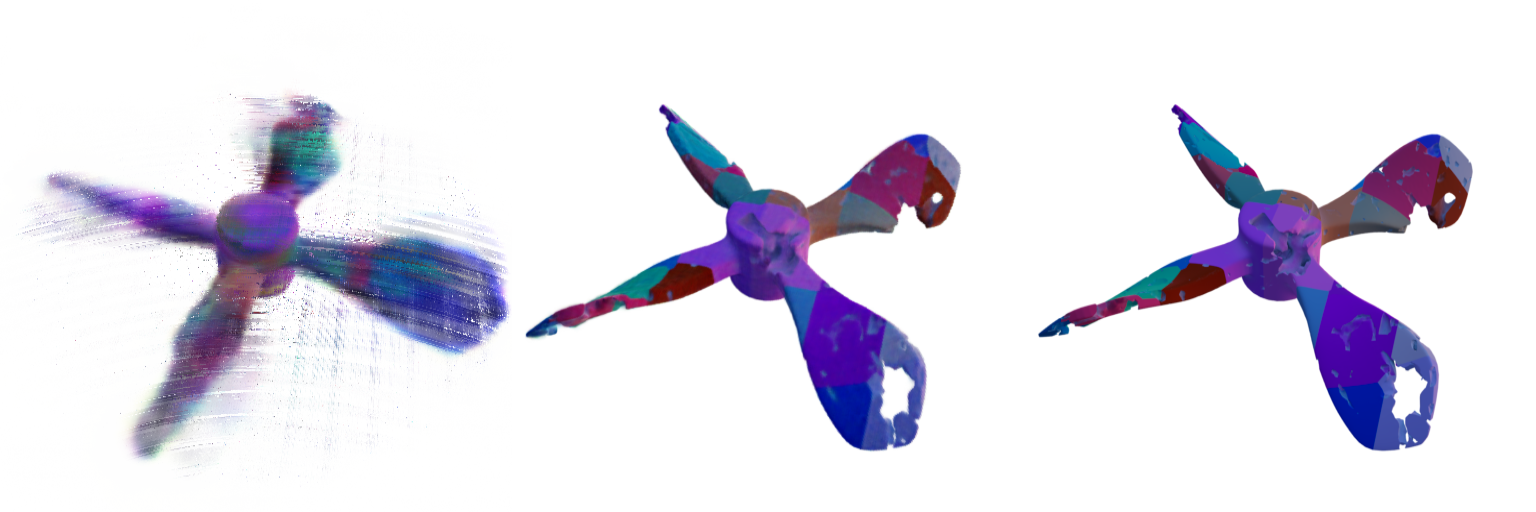}
    \caption{Extrapolation: rotation}
    \end{subfigure}
  \caption{Qualitative comparison on the 'propeller' dataset. These test results show that D-NerF~\cite{dNeRF} can handle the interpolation of compressing well (shown in a) but not extrapolation (b), and fails to handle rotations both interpolation (c) and extrapolation (d), while ours can handle all well.}
  \label{fig:qualitative_synthetics}
\end{figure}

We train D-NeRF \cite{dNeRF} on this dataset with access to every other temporal frame for the interpolation test, and access to the first half of the frames for the extrapolation test. For our model, we train only on the first frame for which we supply a coarse tetrahedral mesh, which can be considered the 'rest` state. The base tetrahedral geometry is trivial to extract from NeRF's density field as described in Section~\ref{sec:ext_generic_objects} and its animation can be produced in off-the-shelf software or programatically. The coarse tetrahedral mesh is illustrated in Figure~\ref{fig:propeller_tet_cage} in its rest state on the left, and rotated by $45$ degrees on the right.
We report the mean PSNR and LPIPS on both interpolation of every other frame (unseen in training) and extrapolation over time (second half of the frames, unseen in training) in Tables~\ref{tab:propeller_quantitative_spin} and~\ref{tab:propeller_quantitative_compress}. Qualitative results are shown in Figure~\ref{fig:qualitative_synthetics}.

Despite having access to significantly more data and using positional encoding on the time signal, the D-NeRF baseline is unable to capture the rotational motion at all (Table~\ref{tab:propeller_quantitative_spin}), and fails to extrapolate on the compression as well (Table~\ref{tab:propeller_quantitative_compress}). In contrast, our method handles both compression and rotation well, albeit with baked-in illumination from the training frame. Using FastNeRF as the rendering method, \methodname{} is also substantially faster, producing images at around $10$ms a frame with resolution $512 \times 512$, as opposed to seconds for the baseline (these times do not include the posing of the tetrahedral mesh which is done offline for this set of experiments, see Section \ref{sec:performance} for a more detailed performance evaluation). Such speeds are possible as MLP evaluations are avoided entirely when FastNeRF and \methodname{} are used in combination. 

\begin{table}[]
    \centering 
        \begin{tabular}{c|cc|cc|}
        \multirow{3}{*}{Method} & \multicolumn{4}{c|}{Spinning}\\
        \cline{2-5}
        {} & \multicolumn{2}{c|}{Interpolation} & \multicolumn{2}{c|}{Extrapolation}\\
        \cline{2-5}
        {} & PSNR $\uparrow$ &  LPIPS $\downarrow$ & PSNR $\uparrow$ & LPIPS $\downarrow$\\
        \hline
        D-NerF & 16.63 & 0.265 & 12.78 & 0.346\\
        Ours & \textbf{27.72} & \textbf{0.022} & \textbf{29.87} & \textbf{0.014} \\
    \end{tabular}
        \caption{\ready{} Quantitative evaluation on the propeller undergoing a spinning motion. While our system performs equally well for interpolation and extrapolation, D-NeRF~\cite{dNeRF} fails to learn a model that is able to handle rotational motion.}
    \label{tab:propeller_quantitative_spin}
\end{table}

\begin{table}[]
    \centering 
        \begin{tabular}{c|cc|cc|}
        \multirow{3}{*}{Method} & \multicolumn{4}{c|}{Compression}\\
        \cline{2-5}
        {} & \multicolumn{2}{c|}{Interpolation} & \multicolumn{2}{c|}{Extrapolation}\\
        \cline{2-5}
        {} & PSNR $\uparrow$ &  LPIPS $\downarrow$ & PSNR $\uparrow$ & LPIPS $\downarrow$\\
        \hline
        D-NerF & \textbf{34.22} & \textbf{0.004} & 17.87 & 0.108\\
        Ours & 31.68 & 0.007 & \textbf{29.83} & \textbf{0.009} \\
    \end{tabular}
        \caption{\ready{} Quantitative evaluation on the propeller undergoing compression. It is interesting to note that unlike our system, the baseline is trained on multiple temporal frames and hence observed appearance changes due to \eg{} change in relative orientation of the lighting with respect to the object. This extra supervisory signal allows the baseline to interpolate a bit better than our system. When it comes to extrapolating, the baseline fails where our system maintains a similar level of quality compared to the interpolation regime.
        }
    \label{tab:propeller_quantitative_compress}
\end{table}

\subsubsection{Generalisation to other Types of Deformation}
\begin{figure}[h]
  \centering
  \includegraphics[width=0.5\textwidth]{{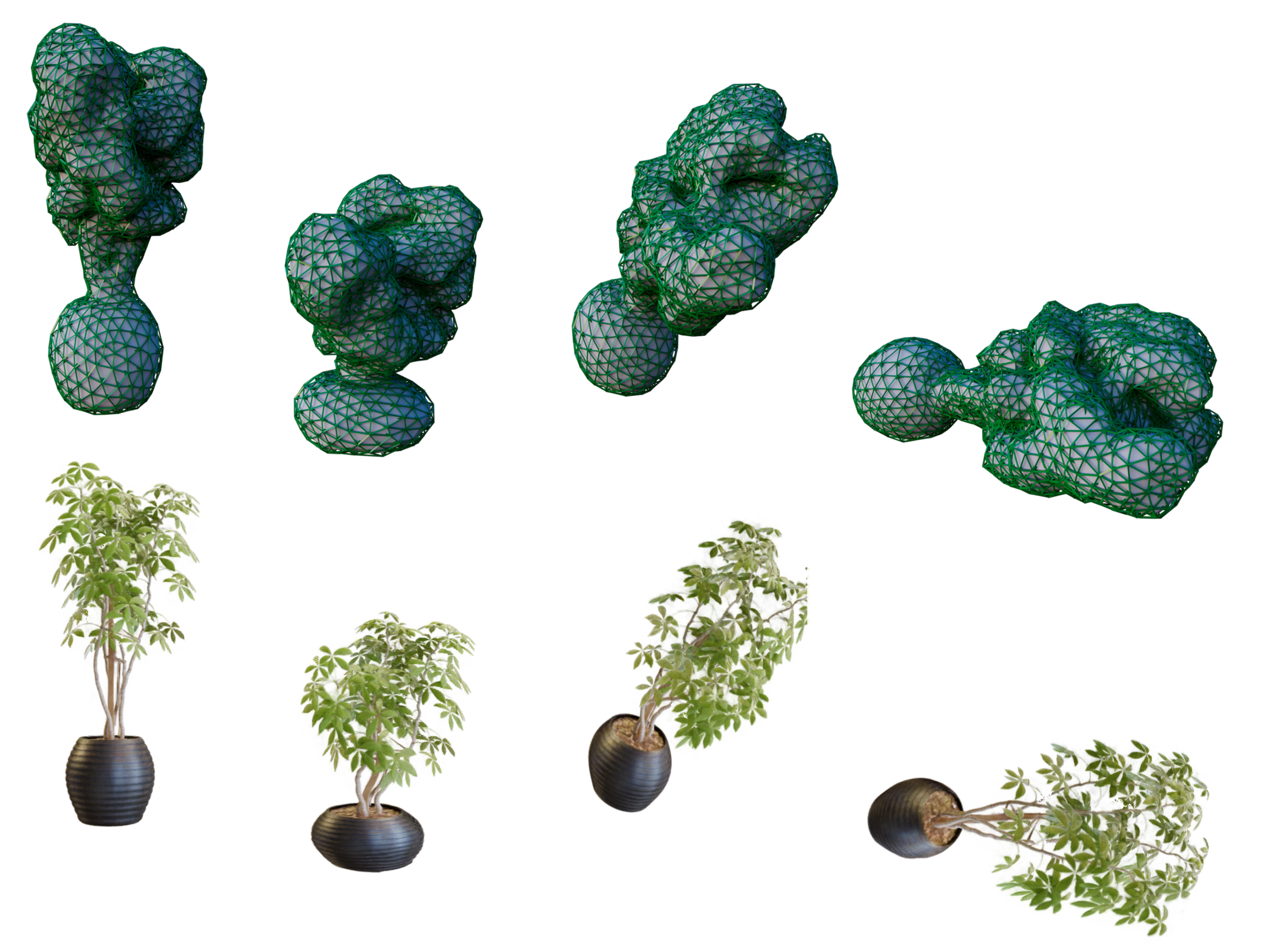}}
  \caption{The NeRF \textbf{ficus} undergoing elastic deformation. \textit{Top row}: Simplified view of the tetrahedral cage surrounding the density mesh extracted from the static scene. \textit{Bottom row}: Rendered result. Best viewed zoomed in.}
  \label{fig:elasticDeformation}

\end{figure}

\begin{figure}[h]
  \centering
  \includegraphics[width=0.5\textwidth]{{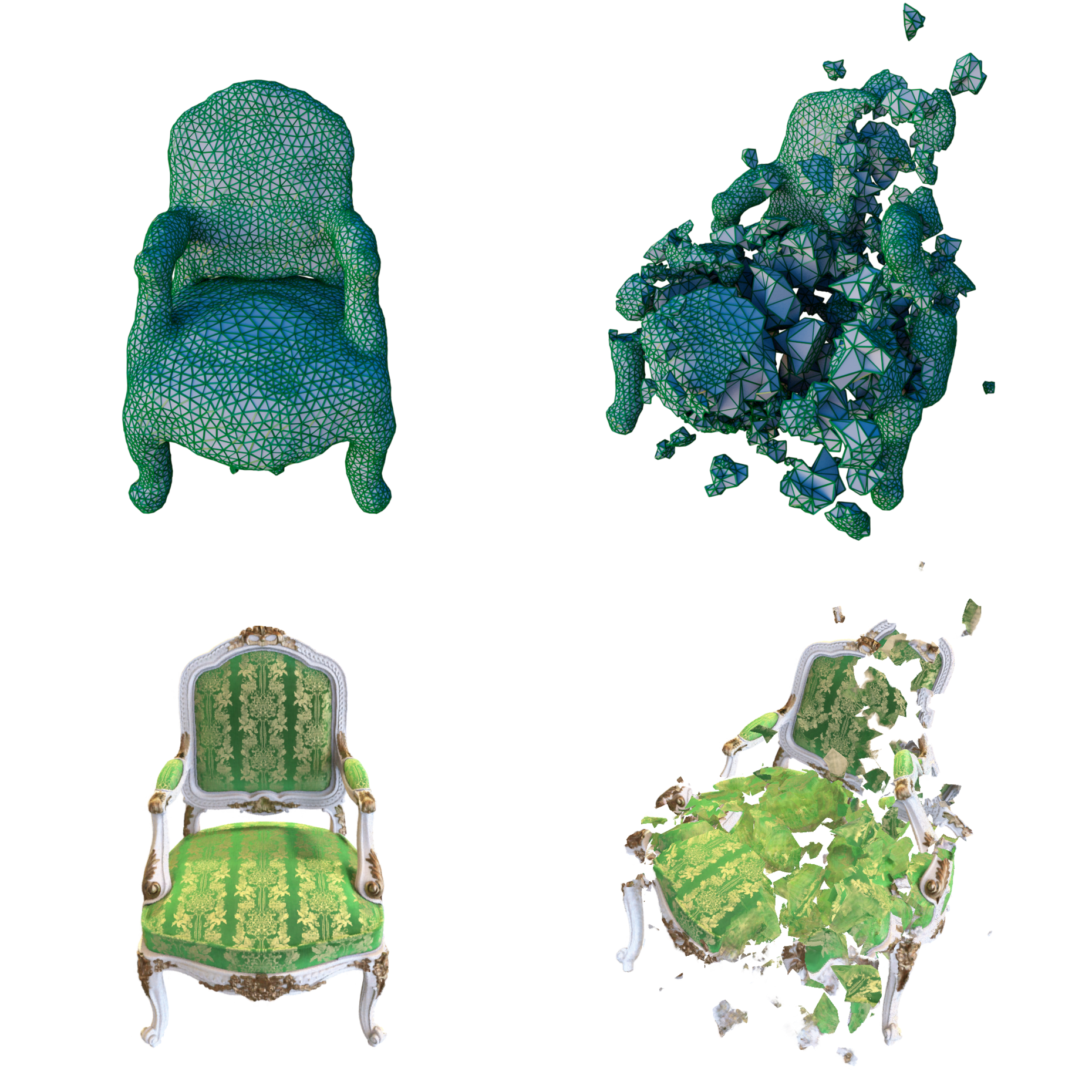}}
  \caption{The NeRF \textbf{chair} shattering. \textit{Top row}: Simplified view of the tetrahedral cage surrounding the density mesh extracted from the static scene. \textit{Bottom row}: Rendered result, note that the mesh is larger than the object contained within it. Best viewed zoomed in.}
  \label{fig:shatterDeformation}
\end{figure}

One of the strengths of \methodname{} is that once a static model is trained, we can generalise to any geometric deformation that can be expressed with the tetrahedral field constructed from its density. This opens new possibilities to use volumetric models in games or AR/VR contexts where a user's manipulation of the environment is not known a priori. To illustrate the generalisation power of \methodname{}, we use the Finite Element Method to apply physical simulations to the tetrahedral mesh and thereby the scene. This allows us to demonstrate effects that have previously eluded volumetric models, such as elastic deformation (Figure~\ref{fig:elasticDeformation}) and shattering (Figure~\ref{fig:shatterDeformation}).

\subsection{Face Data}
\label{sec:face_data}
\ready{}
Our multi-view face data is acquired with a camera rig that captures synchronized videos from 31 cameras at 30~fps. These cameras are located 0.75--1~m from the subject, with viewpoints spanning 270$^{\circ}$ around their head and focusing mostly on frontal views within $\pm 60^{\circ}$. Illumination is not uniform. All the images are down-sampled to 512$\times$512 pixels and colour corrected to have consistent colour features across cameras. We estimate camera poses and intrinsic parameters with a standard structure-from-motion pipeline by Agisoft Metashape~\textregistered \footnote{https://www.agisoft.com/}.

For the experiments, we captured speech sequences with natural head motion for four subjects. Half the subjects additionally performed various facial expressions and head rotations. To train the models for each subject we use the face tracking result from the multi-view system described in ~\cite{wft} and images from multiple cameras at a single time-instance (frame). The frame is chosen to satisfy the following criteria: 1) a significant area of the teeth is visible and the bottom of the upper teeth is above the top of the lower teeth to place a plane between them, 2) the subject looks forward and some of the eye white is visible on both sides of the iris, 3) the face model fit for the frame is accurate, 4) the texture of the face is not too wrinkled (\eg in the nasolabial fold) due to the mouth opening. The last point is important because our system does not account for expression-dependant appearance changes, and we cannot train it with a frame where the mouth opening creates strong appearance changes that are only visible for certain expressions. When a single frame satisfying constraints 1--4 is not available, we use two frames: a frame where the user has a neutral expression looking forward that satisfies 2--4 to train everything but the mouth interior, and a frame with the mouth open and that satisfies 1 and 3 to train the mouth interior.

To evaluate the ability of the baselines and our model to generate novel views, we exclude views from two cameras from training and reserve them for testing. To evaluate the ability of the models to animate the avatars, we use 600 consecutive frames from the captures that exclude training and use them as ground truth for quantitative metrics. Additionally, we evaluate qualitatively an end-to-end system with videos of two of the subjects talking in front of a webcam and performing various facial expressions. We use the real-time monocular face tracker system described in~\cite{wft} to track the users' face, and animate their avatars in a virtual presence application.

\subsection{Qualitative results on faces}
\label{sec:qualitative}
\ready{}
\begin{figure*}
  \centering
  \includegraphics[width=0.9\textwidth]{{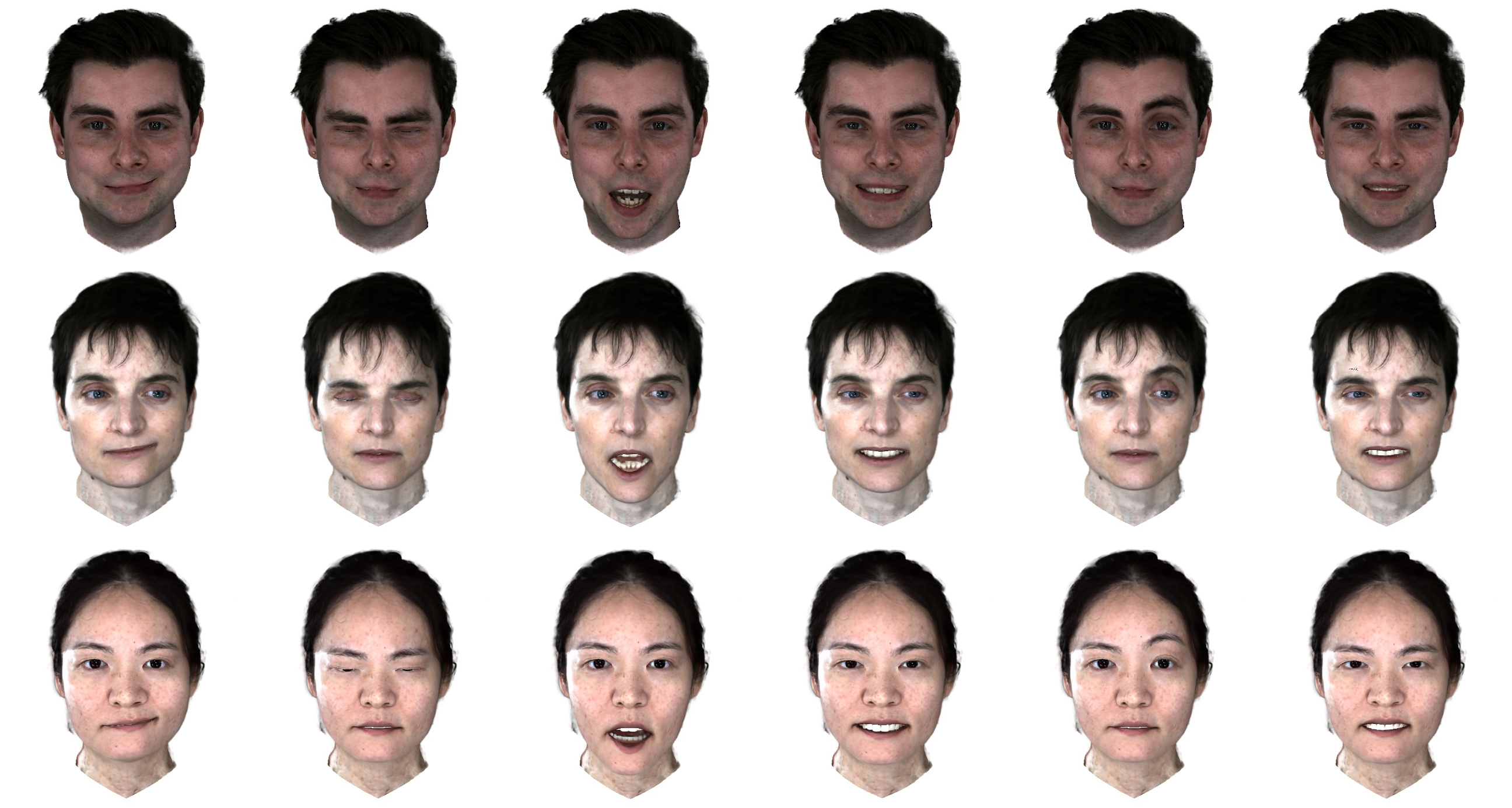}}
  \caption{\ready{} Avatars of three subjects rendered from viewpoints not seen at training time. Each column shows activation of a single facial expression blendshape with all the other blendshape values set to 0. Note that each expression is consistent for all three identities.}
  \label{fig:expr_variety}
\end{figure*}

\begin{figure}
  \centering
  \includegraphics[width=1.0\columnwidth]{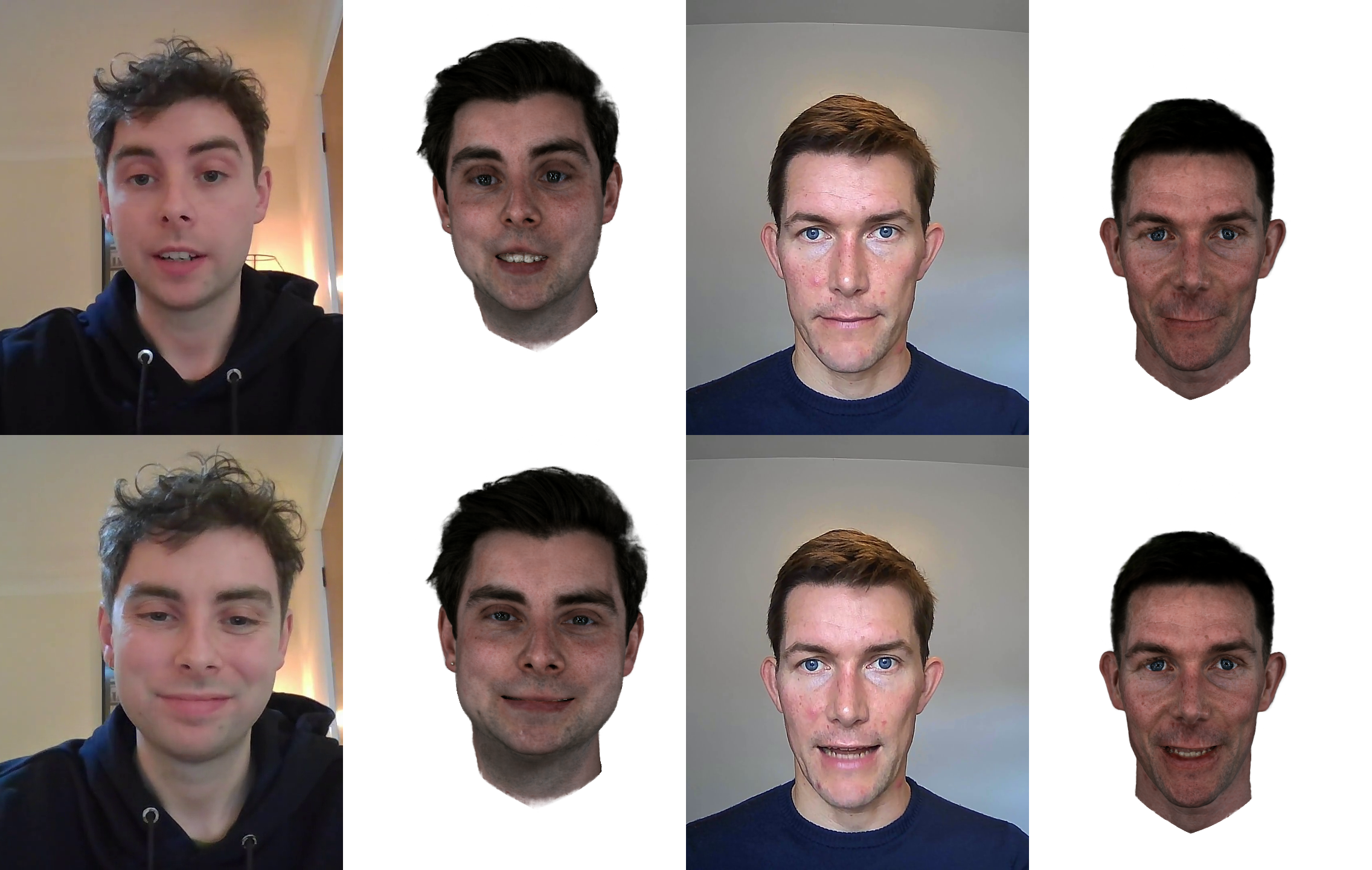}
  \caption{Avatars of two subjects rendered from their webcam recording. In the first part of webcam recordings, each subject speaks and moves their head naturally. In the second half, each subject was asked to enact various facial expressions and head motions.}
  \label{fig:webcam_qualitative}
\end{figure}

We first show qualitative results of face avatars performing various expressions in Figure \ref{fig:expr_variety}. Our method nicely separates the head from the background, without the use of a green screen. Our volumetric 3DMM model allows to seamlessly generate consistent and photo-realistic expressions across different identities. Furthermore, by incorporating an implicit representation of the surface model, we are able to render hair effortlessly. 
The underlying face model allows to synthesize new expressions and poses for each avatar. Our mouth interior model produces plausible images, including expressions with an open mouth and showing teeth (e.g., the third column).  

Being able to render photo-realistic content driven by a self-recorded webcam sequence is an essential step toward creating a telepresence system (scalable rendering). Therefore, we evaluated our method qualitatively with animations driven by webcam recordings. In these recordings, the subject is sitting frontal to the camera (either a laptop camera or an external webcam camera). In the first part of the capture, each subject was asked to vocalise a number of sentences. In the second part, we asked the subjects to enact a couple of facial expressions, head motions and eye movements. The per-frame expression and pose parameters of these sequences are obtained using the monocular face tracker system described in~\cite{wft}, with identity parameters taken (and fixed) from the multi-view fitting result of the training frame (see Section~\ref{sec:face_data}). The resulting avatars from these sequences and their corresponding webcam images are shown side-by-side in Figure \ref{fig:webcam_qualitative}.

\subsection{Comparison to State of the Art}
\ready{}

\begin{figure*}
\setlength{\tabcolsep}{0pt}
\begin{tabular}{ccccccccc}
&\multicolumn{3}{c}{representative renders} &  open mouth & \multicolumn{2}{c}{extreme expressions} & \multicolumn{2}{c}{eye tracking and tongue failures}\\
\begin{turn}{90} \hspace{15pt} ground truth\end{turn} &
  \includegraphics[height=0.16\linewidth]{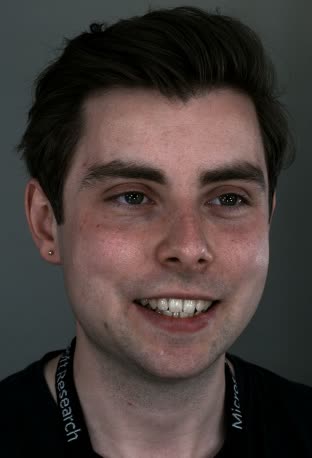}  &
   \includegraphics[height=0.16\linewidth]{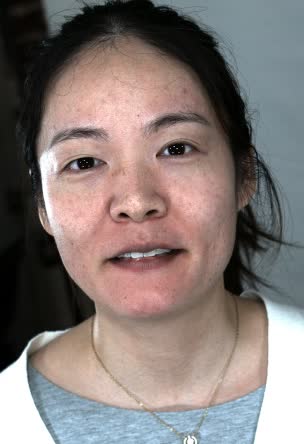} &
   \includegraphics[height=0.16\linewidth]{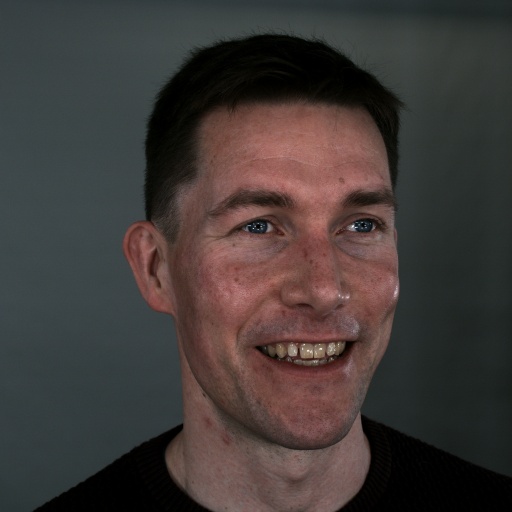}  &
   \includegraphics[height=0.16\linewidth]{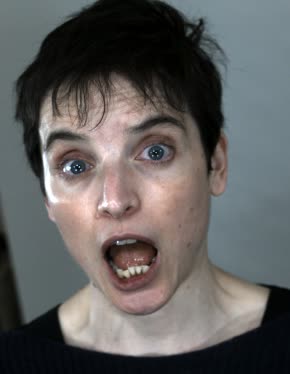} &
   \includegraphics[height=0.16\linewidth]{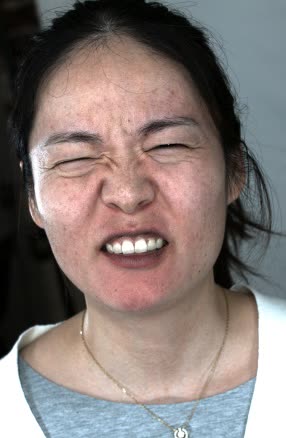} &
   \includegraphics[height=0.16\linewidth]{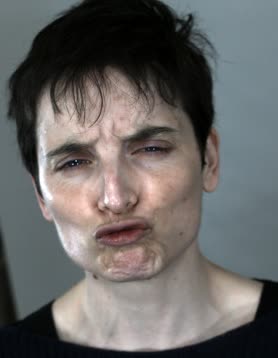} &
   \includegraphics[height=0.16\linewidth]{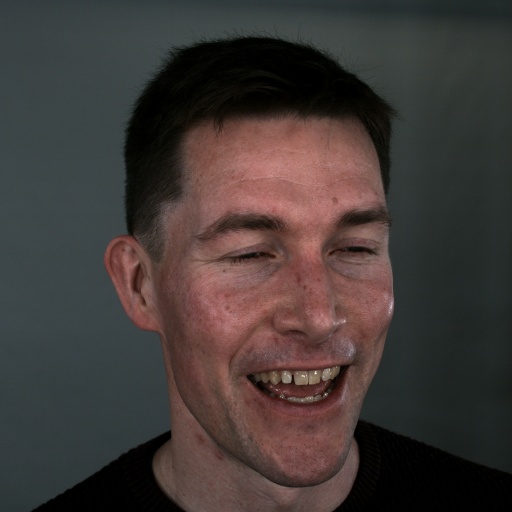} &
   \includegraphics[height=0.16\linewidth]{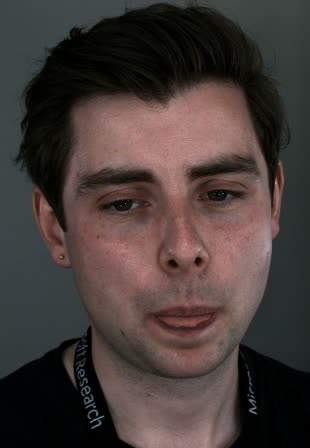} 
   \\
\begin{turn}{90}\cite{neural_head_avatars} \end{turn} &
  \includegraphics[height=0.16\linewidth]{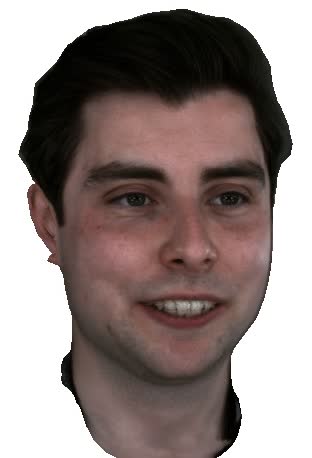}  &
   \includegraphics[height=0.16\linewidth]{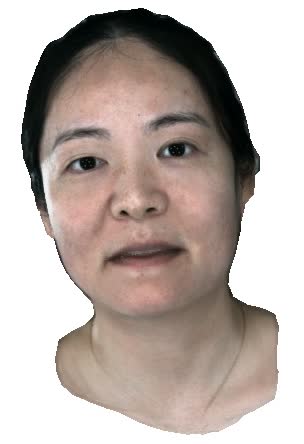} &
  \includegraphics[height=0.16\linewidth]{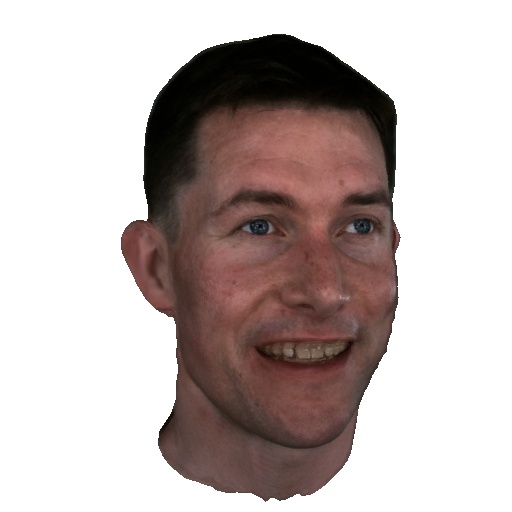}  &
   \includegraphics[height=0.16\linewidth]{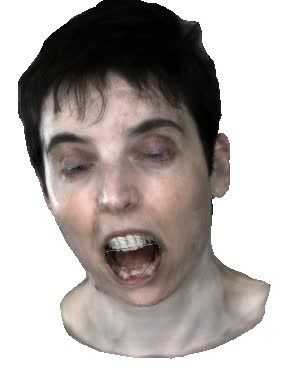}&
   \includegraphics[height=0.16\linewidth]{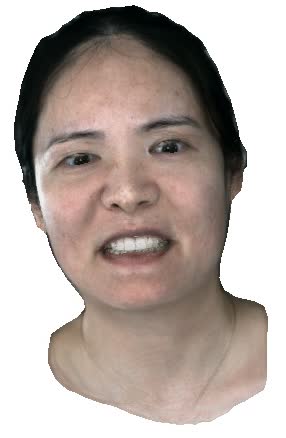}&
   \includegraphics[height=0.16\linewidth]{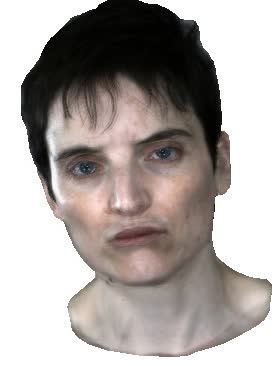}&
  \includegraphics[height=0.16\linewidth]{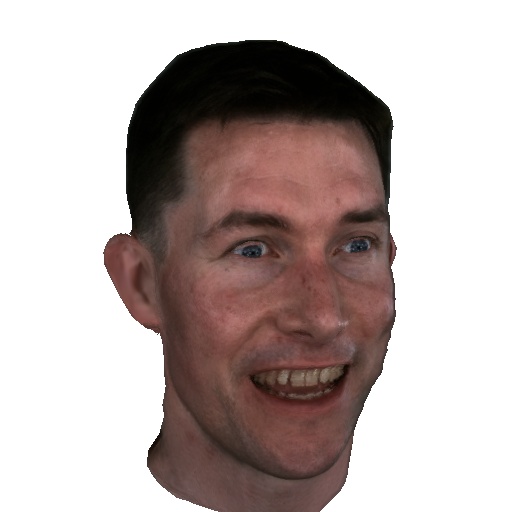}&
   \includegraphics[height=0.16\linewidth]{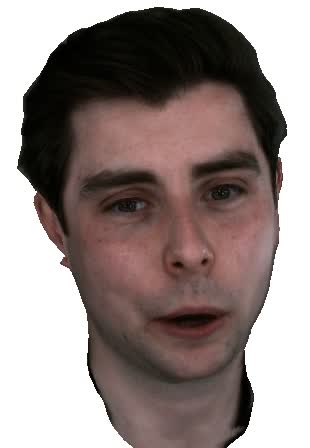}
\\
\begin{turn}{90} \hspace{30pt}ours \end{turn} &
  \includegraphics[height=0.16\linewidth]{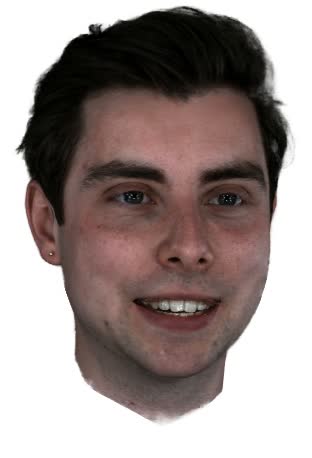}  &
   \includegraphics[height=0.16\linewidth]{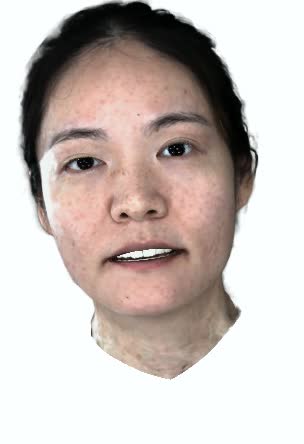} &
  \includegraphics[height=0.16\linewidth]{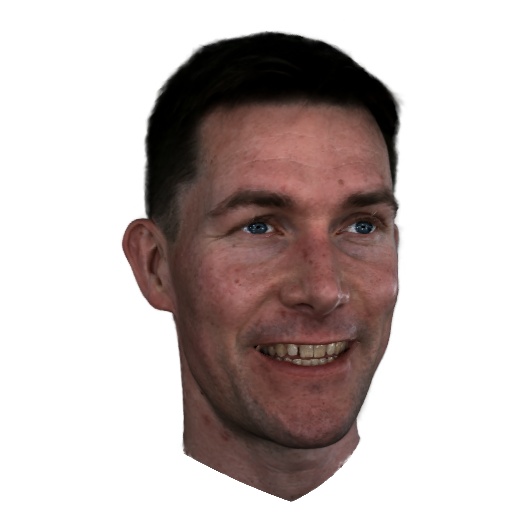}  &
   \includegraphics[height=0.16\linewidth]{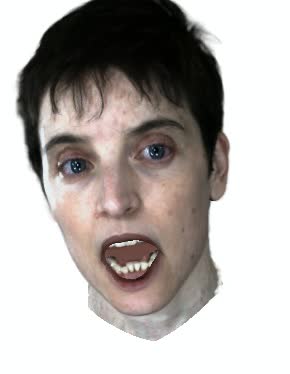}&
   \includegraphics[height=0.16\linewidth]{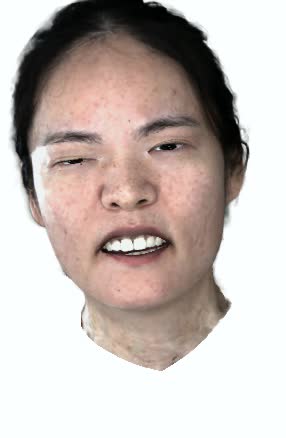} &
   \includegraphics[height=0.16\linewidth]{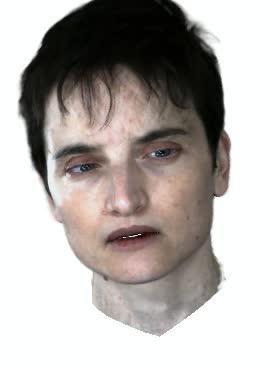}&
   \includegraphics[height=0.16\linewidth]{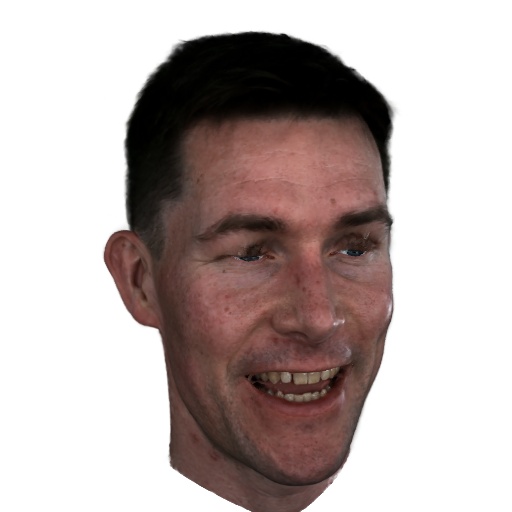}&
   \includegraphics[height=0.16\linewidth]{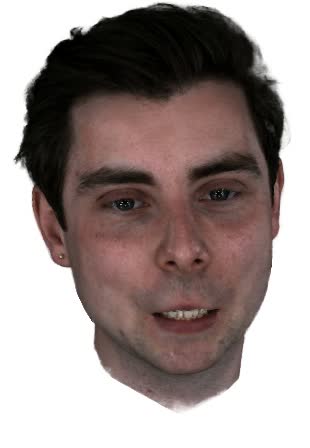}
\end{tabular}
\caption{\ready{} Comparison to Neural Head Avatars~\cite{neural_head_avatars}, a method that learns a deformation and texture on top of a 3DMM face model to animate avatars. The first 3 columns show renders of representative quality for both methods. Unlike our method, Neural Head Avatars learns a deformation on top of a mesh and has difficulties representing structures that can open, close, or fold like the mouth or eyelids (columns 4 and 5) or fine details like the hair and the ears. Both methods rely on an accurate face tracker and a 3DMM to deform the avatar, this leads to limitations when the face model cannot capture extreme expressions or deformations of the tongue (columns 5--8). To focus the comparison on animation, the renders use the training camera used for training the neural head avatars baseline.}
\label{fig:comparison_to_nha}
\end{figure*}

Our baseline method for quantitative evaluation on faces is Neural Head Avatars \cite{neural_head_avatars}, a method that learns a deformation and a texture on top of the FLAME~\cite{flame_citation} model from RGB videos. This 3DMM model defines a mesh of the face that is controlled by identity, pose, and expression parameters estimated by a face tracker. As the mesh only describes the coarse geometry of the face, it is refined by subdivision and deformed with a learned per-vertex deformation to create a finer mesh where the learned texture is applied. To train \cite{neural_head_avatars} we use 1000 frames from a frontal camera and test it for both animation and novel-view synthesis on different 600 frames from the same clip. We use the code provided by the authors and adapted the data and training setup to match their input format.

\begin{figure*}
\setlength{\tabcolsep}{0pt}
\begin{tabular}{ccccccc}
& train view & test view & test view & train view & test view & test view \\
\begin{turn}{90}\cite{neural_head_avatars} \end{turn} &
  \includegraphics[height=0.16\linewidth]{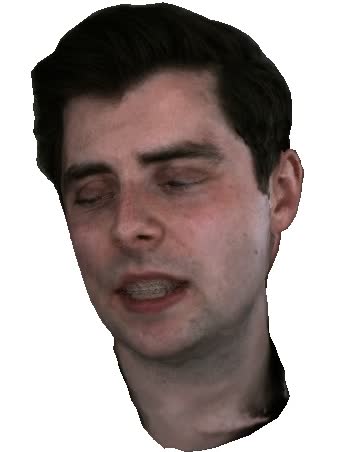}&
  \includegraphics[height=0.16\linewidth]{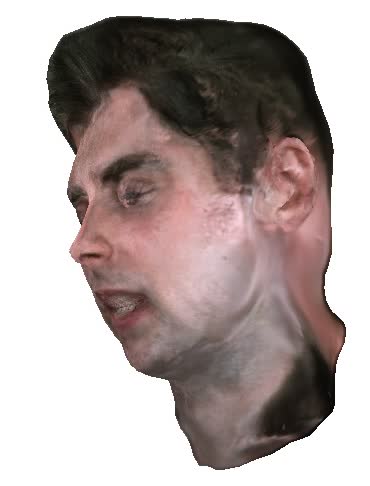}&
  \includegraphics[height=0.16\linewidth]{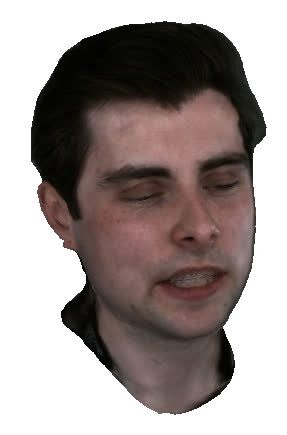} &
   \includegraphics[height=0.16\linewidth]{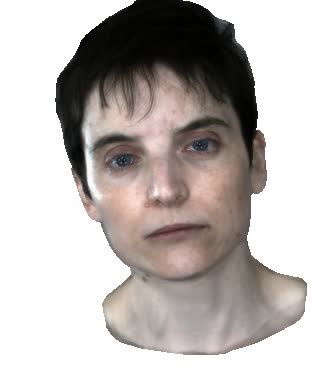}&
  \includegraphics[height=0.16\linewidth]{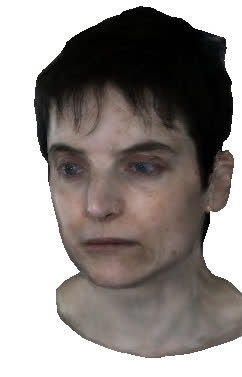}&
  \includegraphics[height=0.16\linewidth]{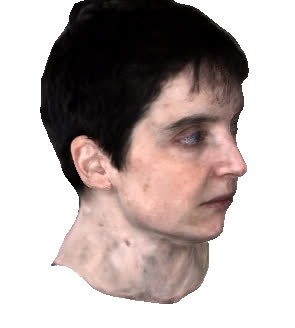}
  \\
  \begin{turn}{90}\hspace{30pt}ours\end{turn} &
  \includegraphics[height=0.16\linewidth]{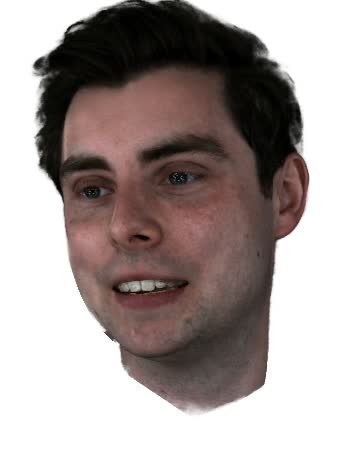} &
  \includegraphics[height=0.16\linewidth]{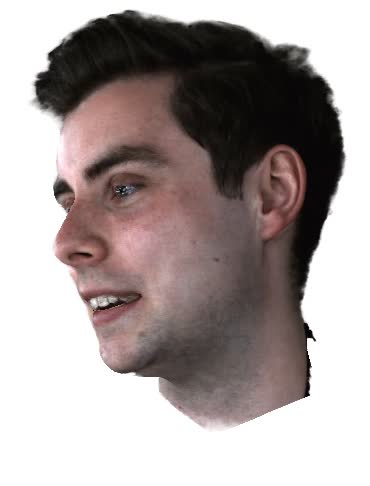} &
  \includegraphics[height=0.16\linewidth]{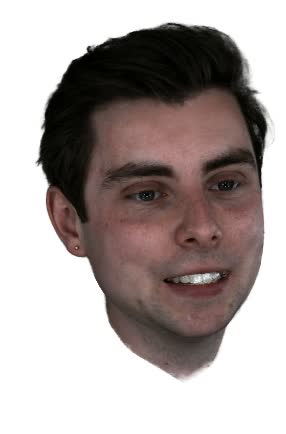} &
  \includegraphics[height=0.16\linewidth]{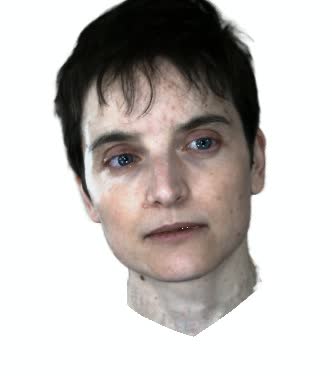}&
  \includegraphics[height=0.16\linewidth]{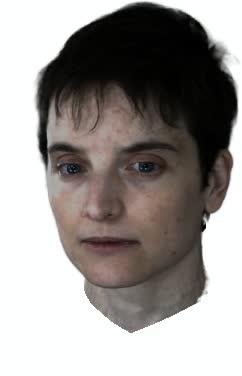}&
  \includegraphics[height=0.16\linewidth]{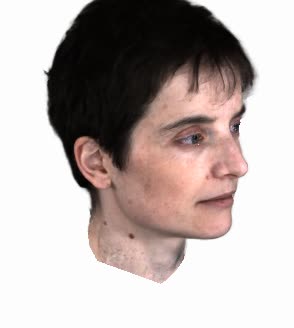}
\end{tabular}
  \caption{\ready{} Comparison to neural head avatars in terms of novel-view rendering. Neural head avatars~\cite{neural_head_avatars} learns a deformation and texture on top of a 3DMM face model from a RGB video with 1000 frames obtained from a single view. Even though the training video contains head rotation, the learned model is highly biased towards the frontal view from the training camera (column 1 and 4) and cannot produce realistic novel views from camera poses noticeably different than the training one (first row, columns 2, 3, and 5, 6). In contrast, our model generalizes well to novel views unseen at training time (second row, columns 2, 3 and 5, 6) because it is trained from a single frame of a multi-camera capture and does not depend on accurate face tracking over time to learn multi-view geometry.}
  \label{fig:comparison_to_nha_testviews}
\end{figure*}

Both our method and \cite{neural_head_avatars} use a 3DMM face model to control face deformations, but we use an implicit representation of the surface while \cite{neural_head_avatars} uses an explicit mesh. As a result, we are able to better represent fine geometry details like hair that is hard to accurately capture using meshes, see Figure~\ref{fig:comparison_to_nha}. Our renders are also sharper because we train the model from a single frame, while \cite{neural_head_avatars} depends on accurate face tracking across multiple frames. Small errors in tracking directly translate to a degradation in the sharpness of the learned textures.

\begin{figure*}
    \centering
    \begin{tabular}{c c@{}c@{}c@{}c@{}c@{}c@{}c@{}c}
        & \multicolumn{4}{c}{Deformation network} & \multicolumn{4}{c}{Ours} \\
        \begin{turn}{90} Opening mouth \end{turn} &
        \includegraphics[width=0.12\linewidth]{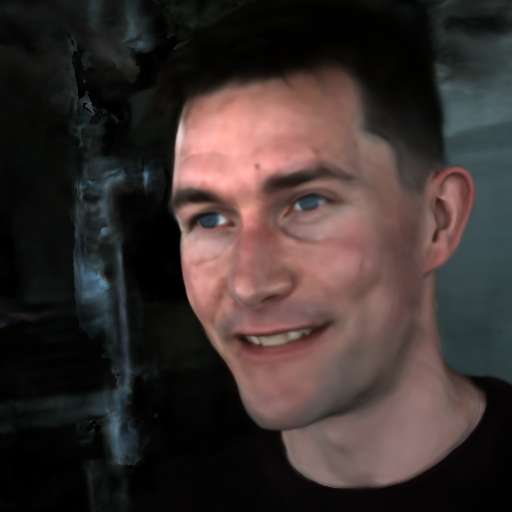} & \includegraphics[width=0.12\linewidth]{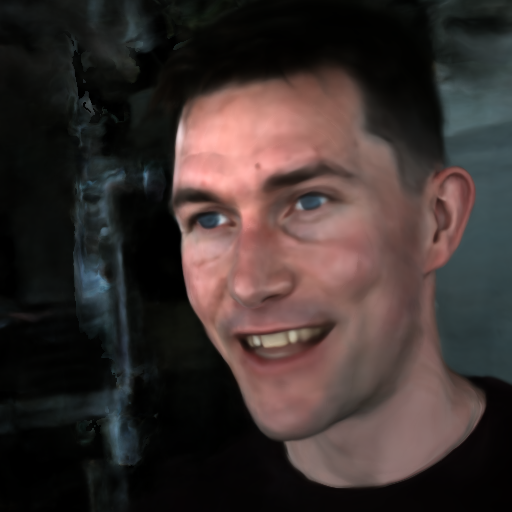} & \includegraphics[width=0.12\linewidth]{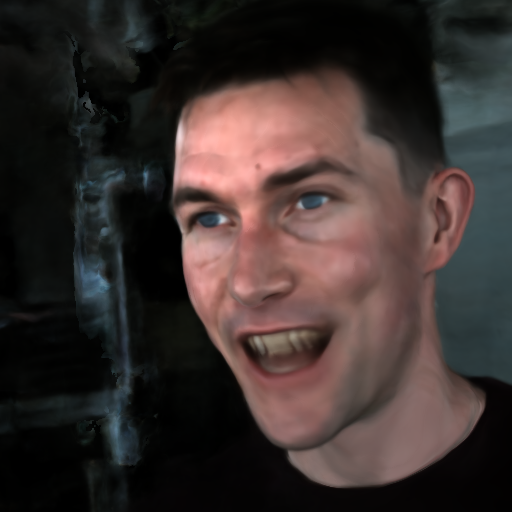} & \includegraphics[width=0.12\linewidth]{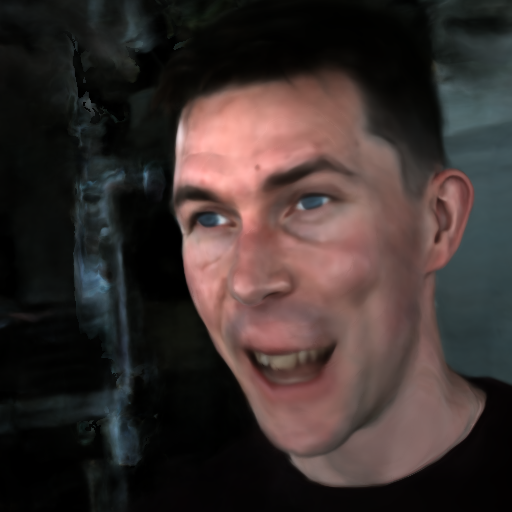} & \includegraphics[width=0.12\linewidth]{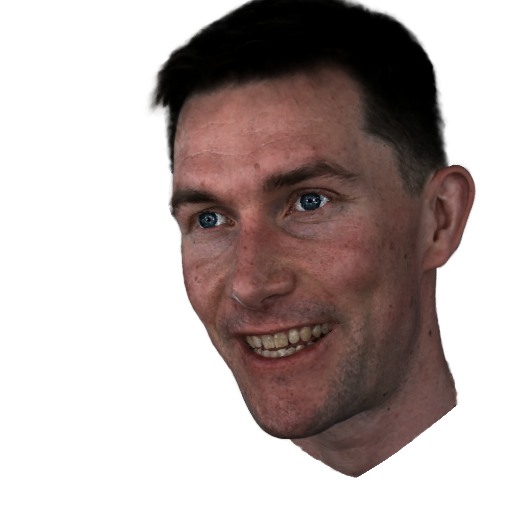} & \includegraphics[width=0.12\linewidth]{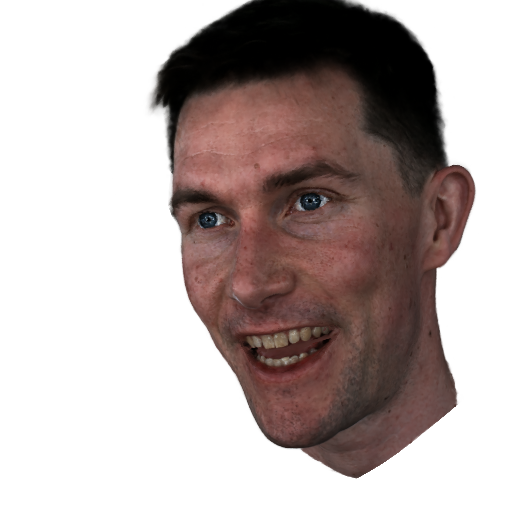} & \includegraphics[width=0.12\linewidth]{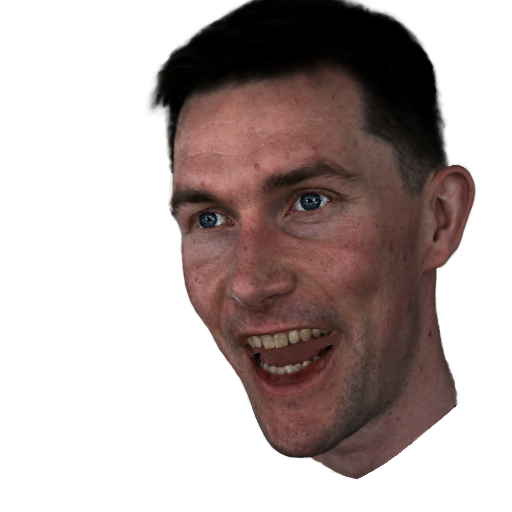} & \includegraphics[width=0.12\linewidth]{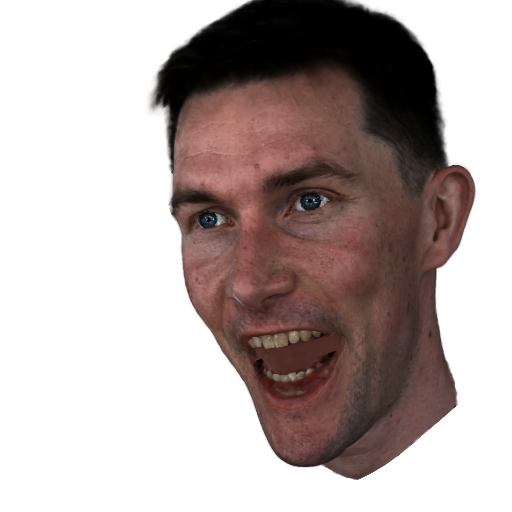} \\
        
        \begin{turn}{90} \quad Clenching \end{turn} &
        \includegraphics[width=0.12\linewidth]{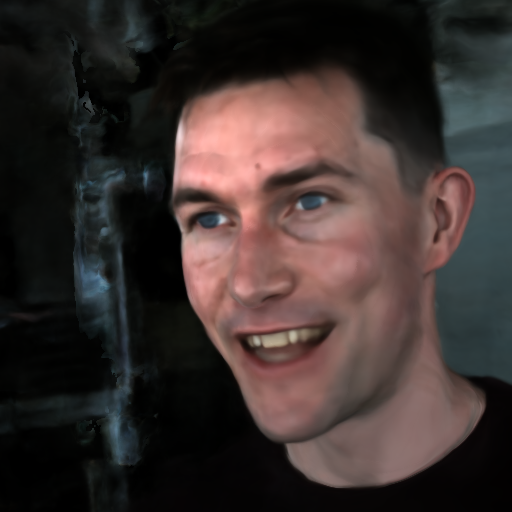} & \includegraphics[width=0.12\linewidth]{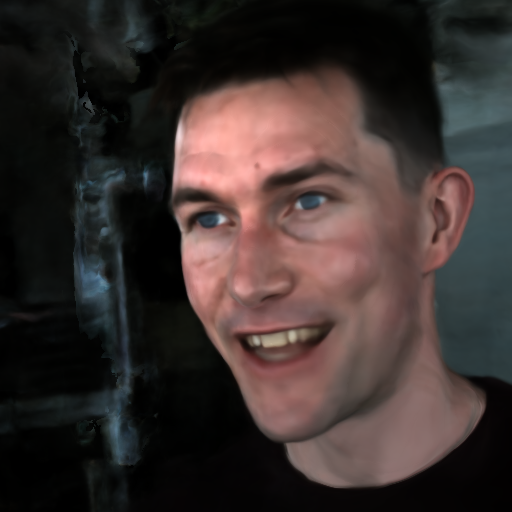} & \includegraphics[width=0.12\linewidth]{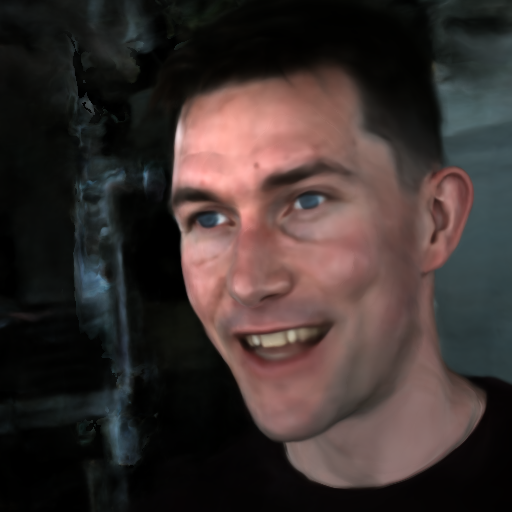} & \includegraphics[width=0.12\linewidth]{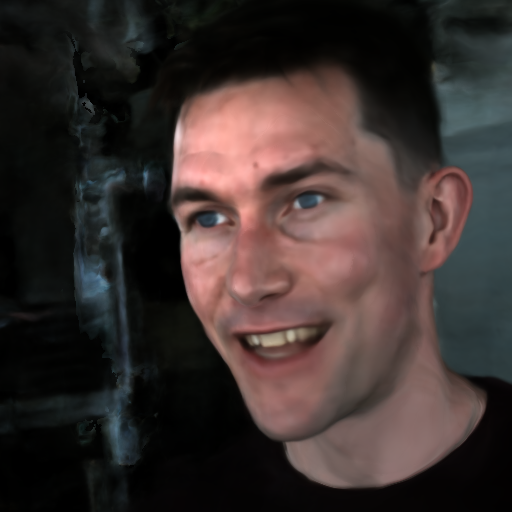} & \includegraphics[width=0.12\linewidth]{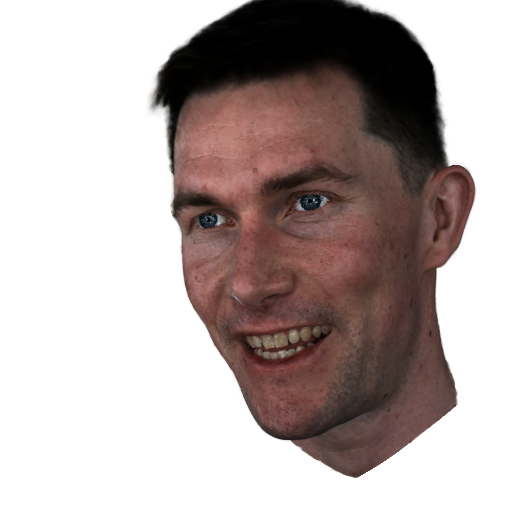} & \includegraphics[width=0.12\linewidth]{images/comparison_deformation/tetnerf_start_frame_0000_expression_005_step_01.png} & \includegraphics[width=0.12\linewidth]{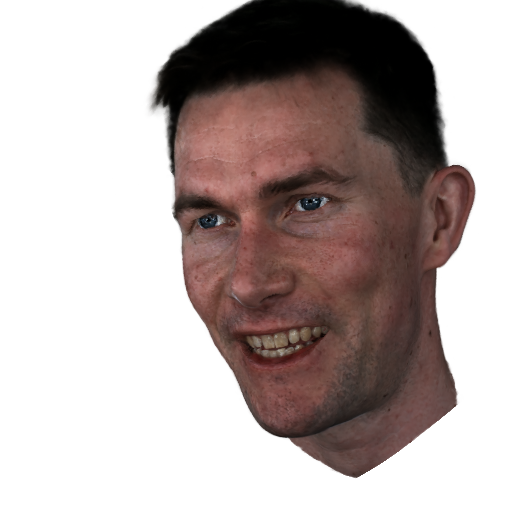} & \includegraphics[width=0.12\linewidth]{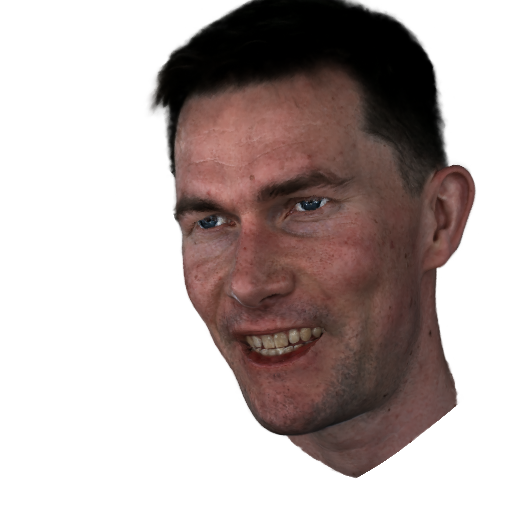} \\
        
        \begin{turn}{90} Cheek raising \end{turn} &
        \includegraphics[width=0.12\linewidth]{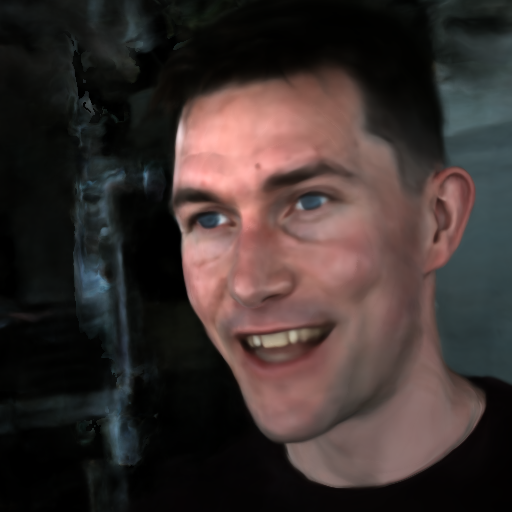} & \includegraphics[width=0.12\linewidth]{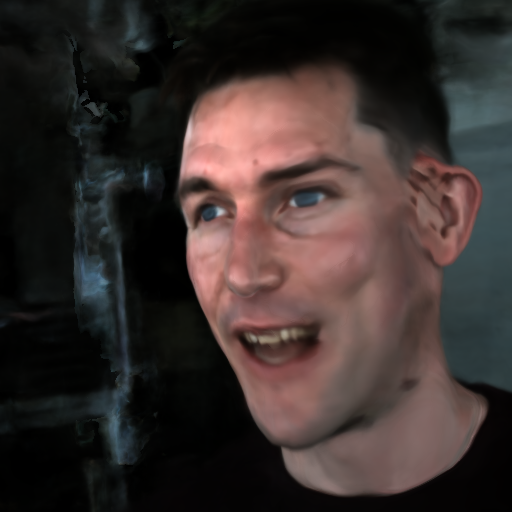} & \includegraphics[width=0.12\linewidth]{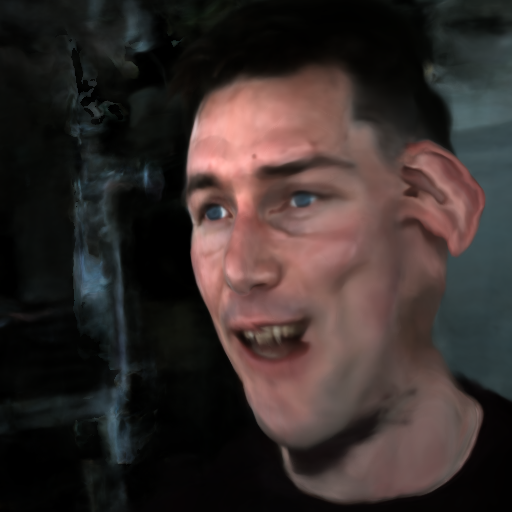} & \includegraphics[width=0.12\linewidth]{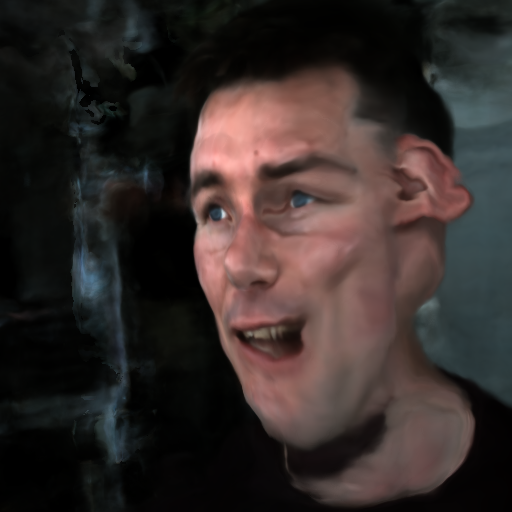} & \includegraphics[width=0.12\linewidth]{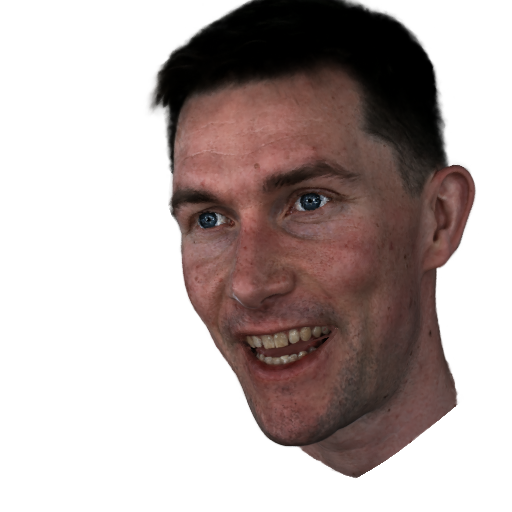} & \includegraphics[width=0.12\linewidth]{images/comparison_deformation/tetnerf_start_frame_0000_expression_009_step_01.png} & \includegraphics[width=0.12\linewidth]{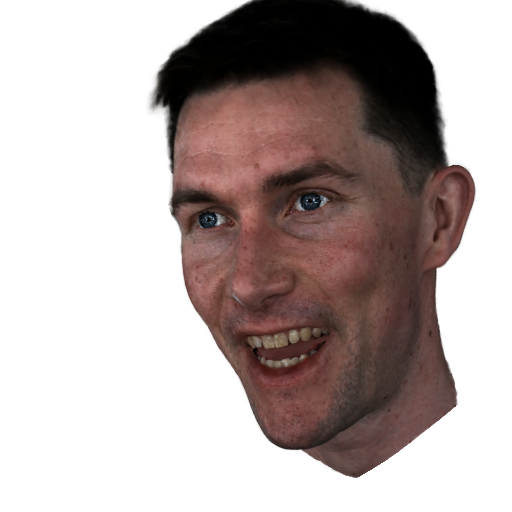} & \includegraphics[width=0.12\linewidth]{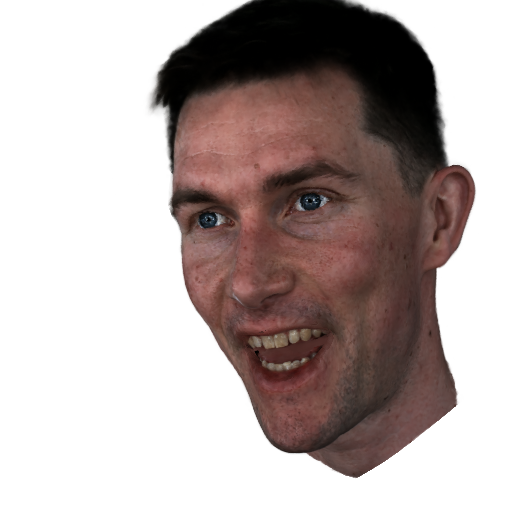} \\
    
    \end{tabular}
    \caption{\ready{} \textbf{Extrapolation using a deformation network} on expressions unseen during training~\eg mouth opening, clenching, and puffing does not yield satisfactory results. The baseline presented here is very similar to \cite{dNeRF}, where we introduce a coordinate-based MLP conditioned on the parameters of the face model to learn a deformation of the samples into canonical space, where a NeRF network is trained. The baseline can only open the mouth as wide as it has seen during training and produce incorrect deformation for unseen expressions (no deformation for clenching, unrelated deformation for cheek raising). Our method, on the contrary, can open the mouth beyond the examples seen in training and generalize to unseen expressions by leveraging the deformations defined by the Vol3DMM face model.}
    \label{fig:extrapolation_deformation_baseline}
\end{figure*}

Figure~\ref{fig:comparison_to_nha} evaluates the avatar animation from the training view: our avatars look more realistic across different subjects and expressions, and our special treatment of the mouth interior produces plausible results for expressions with the mouth wide open where \cite{neural_head_avatars} faces significant quality degradations. Both methods share limitations caused by using the parameters of a 3DMM model to control deformations and a face tracker to estimate those parameters: visual quality degrades when parts of the images are not modelled by the face model,~\eg when the user's tongue becomes visible, or when the face tracker does not accurately track extreme or small expressions like eye closures.

\begin{table}[]
    \centering
    \begin{tabular}{c|c|c}
        Method & PSNR $\uparrow$ & LPIPS $\downarrow$ \\
        \hline
        NHA \cite{neural_head_avatars} & 30.06 & 0.032 \\
        Ours & \textbf{30.20} & \textbf{0.029} \\
    \end{tabular}
    \caption{\ready{} Mean PSNR and LPIPS for all the subjects in the face dataset. The metrics are computed inside a mask derived from foreground/background segmentation masks extracted from each method individually. }
    \label{tab:quantitative_comparison}
\end{table}

We evaluate quantitatively the animation ability of both methods by measuring the PSNR and LPIPS to ground truth images over the head region and report results in Table~\ref{tab:quantitative_comparison}. Our method has better PSNR than \cite{neural_head_avatars} by about 0.1dB and offers a 10\% improvement in LPIPS. This matches with the qualitative results of Figure~\ref{fig:comparison_to_nha}: while a pixel-to-pixel comparison like PSNR shows a moderate improvement over the baseline, a metric measuring patch statistics and able to measure sharpness like LPIPS shows a marked improvement and captures human perception.

Figure~\ref{fig:comparison_to_nha_testviews} shows the ability of the models to generate realistic novel views for two different subjects, the first one trained on a clip with natural speech and the second one trained on a clip with head rotation and speech. The ability of \cite{neural_head_avatars} to generate novel views depends highly on the amount of head rotation available in the training clip: it obtains realistic and consistent novel views for the second subject trained with head rotation, while it produces unrealistic renders for views little seen at training for the first subject trained with a speech clip. Our method produces good quality renders from novel views because it is trained on multiple view points from a single frame, and the renders are still sharper and more photorealistic.

We further compare our method with models that use conditioning to deform a neural radiance field in Figure \ref{fig:extrapolation_deformation_baseline}. We train a deformable NeRF model similar to \cite{dNeRF} by conditioning an MLP on the parameters of the face model and using it to deform samples from physical space into a canonical pose where a standard NeRF model is trained. We use conditioning with a deformation network, as opposed to directly conditioning NeRF as in \cite{Gafni_2021_CVPR}, because the 3DMM face parameters only describe geometric deformations and have considerably less impact on appearance. In particular, we condition a coordinate-based MLP on the expression and pose parameters of the 3DMM model and added an inductive bias with a regularizer that constrains the canonical pose to a zero deformation that matches the 3DMM structure. Figure \ref{fig:extrapolation_deformation_baseline} compares the ability to extrapolate of our model as compared to the baseline with three different expressions: mouth opening, clenching, and cheek raising. The baseline can only open the mouth as wide as it has seen during training, and does not produce meaningful deformations for the unseen expression of clenching and cheek raising. Our method can open the mouth beyond the examples seen in training and generalise to unseen expressions by leveraging the deformations defined by the 3DMM face model. This highlights the main limitation of methods built on conditioning: they do not generalise beyond expressions seen at training time because they require a lot of samples to cover the high dimensional space of expressions and pose (240 dimensions) to work well. Moreover, generalising to unseen expressions with minimum amount of training data demonstrates scalable enrolment capabilities. None of our captures (1000-2000 frames) was long enough to provide meaningful quantitative metrics for this baseline.

\subsection{Analysis of performance}
\label{sec:performance}
\ready{}
\begin{table}[]
    \centering
    \begin{tabular}{c|c|c|c}
        \multirow{2}{*}{Setting} & \multicolumn{3}{|c}{Rendering time in ms}\\
        \cline{2-4}
        {} & Python time & C++ time & Total time \\
        \hline
        1280$\times$720; cache 1024 & 14.79 & 15.68 & 30.47\\
        1280$\times$720; cache 512 & 14.34 & 12.09 & 26.43\\
        854$\times$480; cache 1024 & 14.40 & 9.39 & 23.79\\
        854$\times$480; cache 512 & 13.84 & 6.90 & 20.74\\
    \end{tabular}
    \caption{\ready{} Rendering times for \methodname{} combined with FastNeRF for a single frame averaged over all frames of a webcam video sequence. All the times are in milliseconds, cache size refers to the position-dependent cache size of FastNeRF, view direction dependent cache size was set to 256 in all cases.}
    \label{tab:performance_evaluation}
\end{table}

One of the key aspects of a scalable rendering method is how fast it can render and whether it reaches a frame rate that allows for easily interacting with the scene, what one could call an "interactive frame rate". In this section, we evaluate the combination of \methodname{} and FastNeRF in this aspect. To do so, we render one of the webcam sequences used in Section \ref{sec:qualitative} under two different resolutions and cache sizes. Cache size is a key aspect of FastNeRF that trades quality and memory for speed, for additional details please see the FastNeRF paper. 

The total renderings times, shown in Table \ref{tab:performance_evaluation}, indicate that \methodname{} can render at over 30 frames per second even at high resolution and with a very large cache, while it can be significantly faster for lower cache sizes and lower resolutions. In fact, we believe that \methodname{} could be significantly faster with an optimised implementation, especially because only a part of our rendering pipeline is currently implemented in CUDA/C++ and a large part of it is unoptimised and implemented in Python. To give an indication of possible performance gains, we report how the total rendering time is split between Python and CUDA/C++ for each scenario in Table \ref{tab:performance_evaluation}. The time spent in Python constitutes 49\% or more of the total rendering time for each setting, indicating that there may be a lot of opportunity to further optimise the code. Moreover, the Python time is quite consistent for all the settings as most of our Python computation is independent of the image size (for example posing of the tetrahedral mesh).

\subsection{Ablation studies}

\begin{figure}
  \centering
  \begin{subfigure}[h]{0.04\columnwidth}
    \begin{turn}{90}
      Na\"{i}ve $\quad \quad$ Ours
    \end{turn}
  \end{subfigure}
  \begin{subfigure}[h]{0.94\columnwidth}
    \includegraphics[width=0.94\columnwidth]{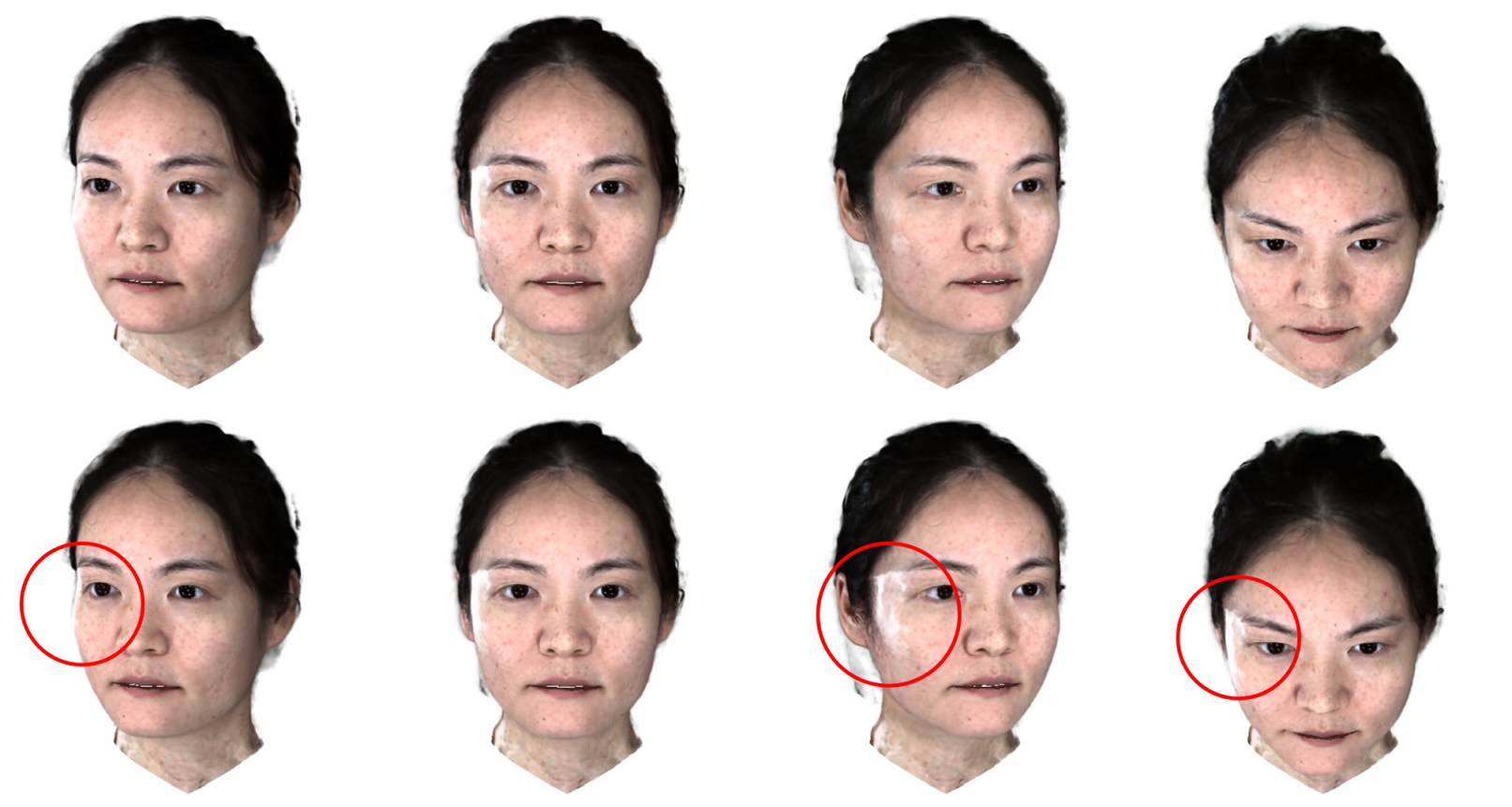}
  \end{subfigure}
  \caption{\ready{} \textbf{View dependent effects.} A face rendered under different head poses in a static camera with view directions rotated by the tetrahedral mesh (top) and with non-rotated view directions (bottom). The face with an approximately neutral head pose is similar across both scenarios as this pose is close to the one seen in the enrolment frame. Other head poses show a fixed position of the specularities when view directions are not rotated based on the motion of the tetrahedral mesh.}
  \label{fig:ablation_view_direction}
\end{figure}

\textbf{Accounting for changes in view direction}
To evaluate the importance of rotating the view directions with the rotation of tetrahedra (Section~\ref{sec:view_dir_rotation}) we show a series of face images rendered with and without this feature in Figure~\ref{fig:ablation_view_direction}. All these images are rendered with the same camera pose and the only varying factor is the pose of the head relative to the neck. The renders where view directions were not rotated show specular reflections that are `attached' to the face surface and move together with the head, which creates uncanny effects, including implausible specular patterns on the side of the head and temple. When we rotate view directions according to the rotation of the tetrahedra, as proposed in section \ref{sec:view_dir_rotation}, the specular reflections are no longer attached to the face surface and the images look more plausible.

\begin{figure}
  \centering
  \begin{subfigure}[h]{0.04\columnwidth}
    \begin{turn}{90}
      Na\"{i}ve $\quad \quad$ Ours
    \end{turn}
  \end{subfigure}
  \begin{subfigure}[h]{0.94\columnwidth}
    \includegraphics[width=0.94\columnwidth]{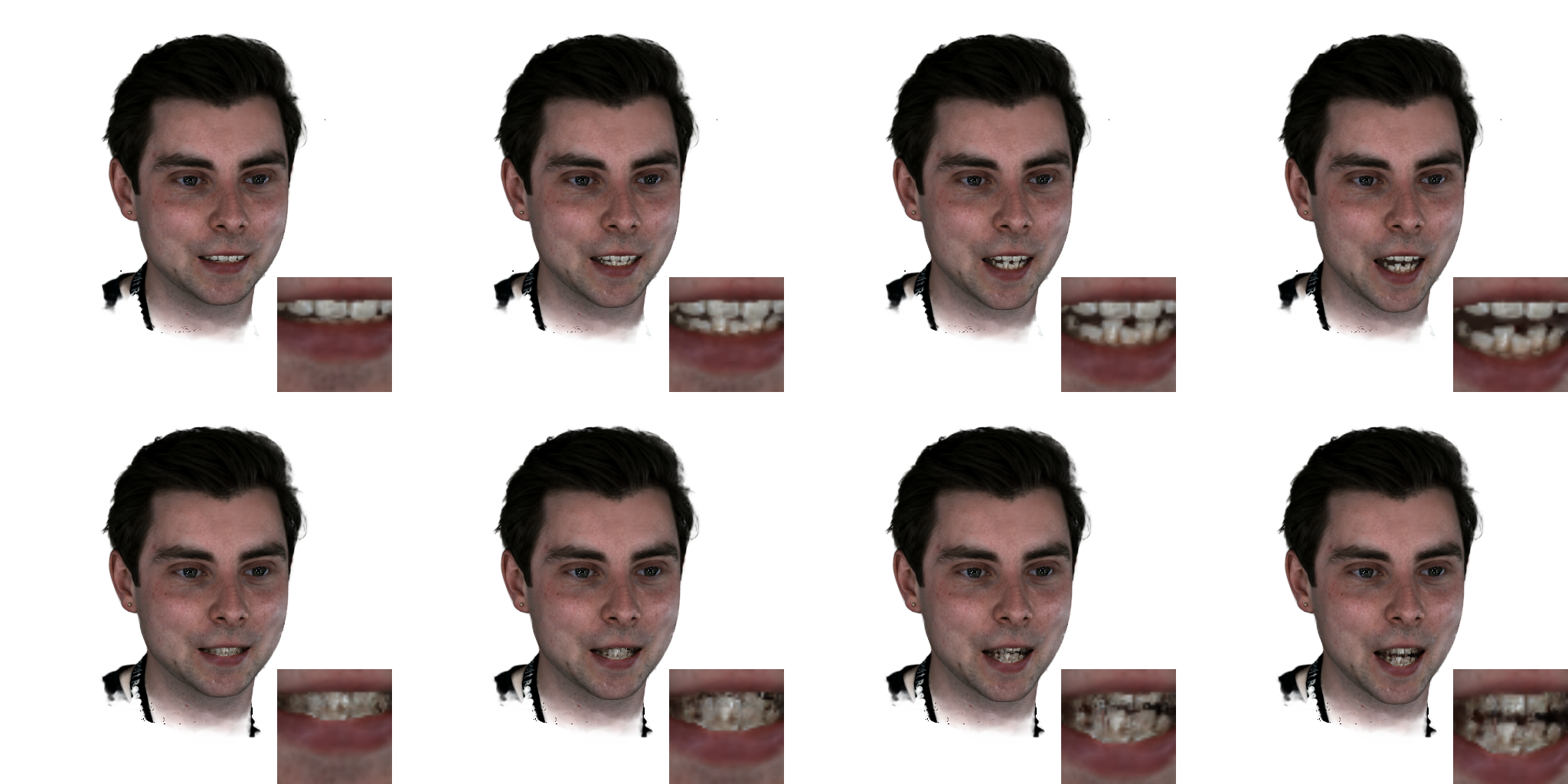}
  \end{subfigure}
  \caption{\ready{} \textbf{Mouth interior.} A face with the mouth progressively opening (left to right) trained and rendered using a geometric two-plane model for the mouth interior (top) and with a naive fully tetrahedral model (bottom). Two-plane model for the mouth interior allows for accurate modelling of the teeth appearance and motion, while the fully tetrahedral approach results in artefacts due to teeth being stretched and compressed.}
  \label{fig:05_experiments/ablation_teeth_model}
\end{figure}
\ready{} \textbf{Geometry model for mouth interior}.

Figure~\ref{fig:05_experiments/ablation_teeth_model} qualitatively compares the appearance of the mouth interior between our geometric model driven by two planes and a model that naively deforms the mouth interior with tetrahedra. A fully tetrahedral model stretches and compresses the teeth, resulting in artefacts and incorrect appearance. In contrast, our geometric model of the mouth interior driven by two planes correctly models the appearance of the teeth as the mouth opens.

\begin{figure}
  \centering
  \begin{subfigure}[h]{0.04\columnwidth}
    \begin{turn}{90}
      w/o sparsity $\quad \quad \;$ Ours
    \end{turn}
  \end{subfigure}
  \begin{subfigure}[h]{0.95\columnwidth}
    \includegraphics[width=0.95\columnwidth]{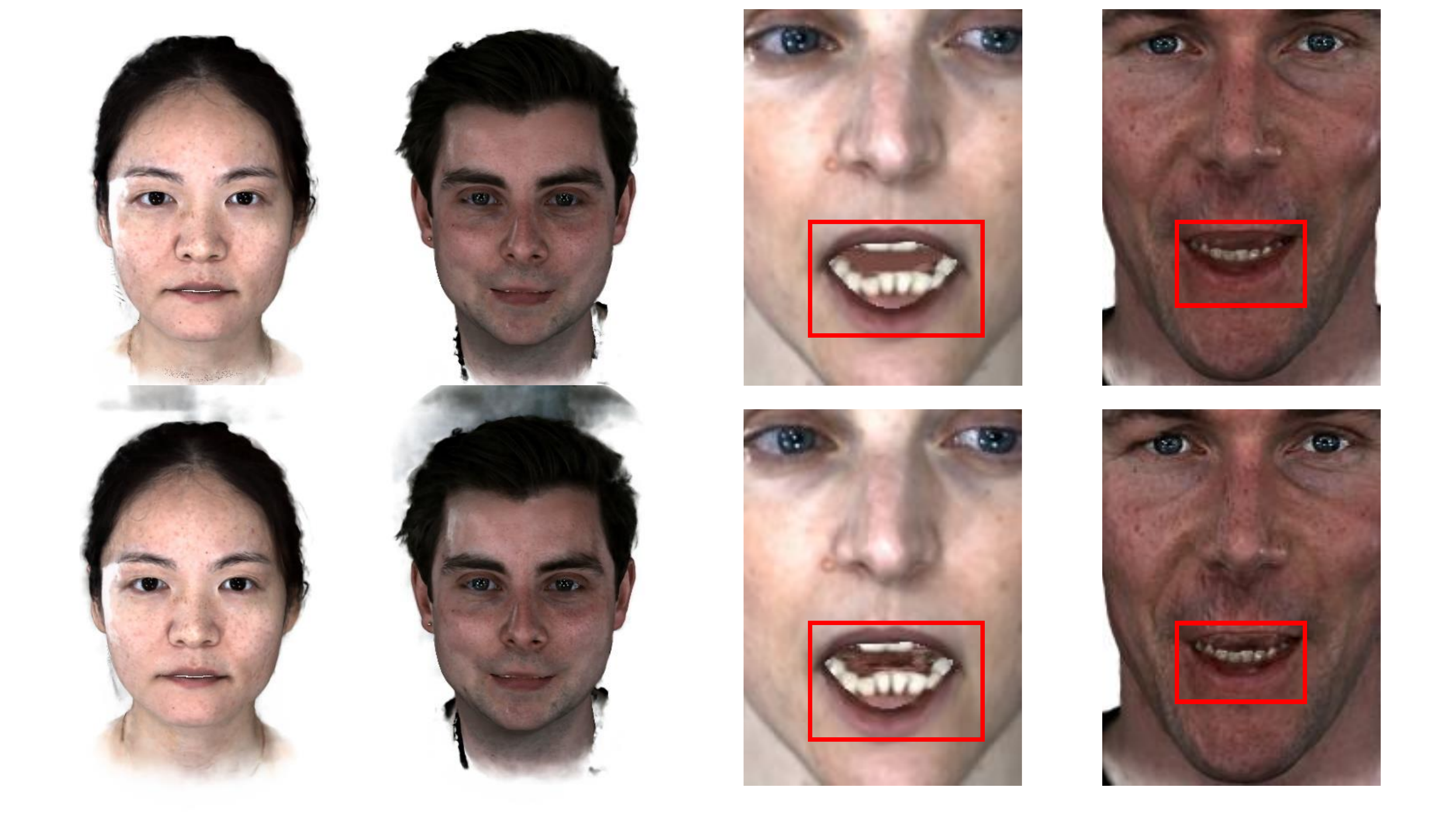}
  \end{subfigure}
  \begin{subfigure}{.45\columnwidth}
  \centering
  \caption{Density around the head}
  \end{subfigure}
  \begin{subfigure}{.53\columnwidth}
  \centering
  \caption{Density in the mouth interior}
  \end{subfigure}
  \caption{\ready{} Four face avatars trained with (top) and without (bottom) sparsity losses. Sparsity losses successfully remove the incorrectly reconstructed background density (first two columns) and spurious density inside the mouth interior visible due to disocclusions (second two columns). Images in (b) best viewed zoomed in.}
  \label{fig:05_experiments/ablation_sparsity}
\end{figure}

\ready{}\textbf{Sparsity}. Figure~\ref{fig:05_experiments/ablation_sparsity} qualitatively compares the impact of our sparsity losses for the space around the head and for the mouth interior. Without a sparsity loss there is density present in the volume surrounding the head because limited training views result in an incorrect reconstruction of the background. Applying a sparsity loss to the background samples that fall in the tetrahedra removes that volume. Similarly, when the mouth is open beyond the training frame, white density appears in disoccluded regions inside the mouth when the sparsity loss is not applied. Applying a sparsity loss to the volume inside the mouth interior results in removal of this spurious density, and allows for controlling the colour observed when the mouth is open beyond the training frame.

\begin{table}[]
    \centering
    \begin{tabular}{c|c|c}
        Method & NeRF Synthetic & Faces \\
        \hline
        Two MLPs & 29.32 & 36.31 \\
        One MLP, NeRF HS & \textbf{29.49} & 36.25 \\
        One MLP, our HS & \textbf{29.48} & \textbf{36.58} 
    \end{tabular}
    \caption{\ready{} Effect of number of networks and hierarchical sampling method on novel-view synthesis of static scenes, as evaluated with Peak Noise-to-Signal Ratio (PSNR) on NeRF synthetic scenes~\cite{nerf} and a single frame from each of the four face datasets.}
    \label{tab:05_experiments/ablations/importance_sampling}
\end{table}

\begin{figure*}
  \centering
  \includegraphics[width=\textwidth]{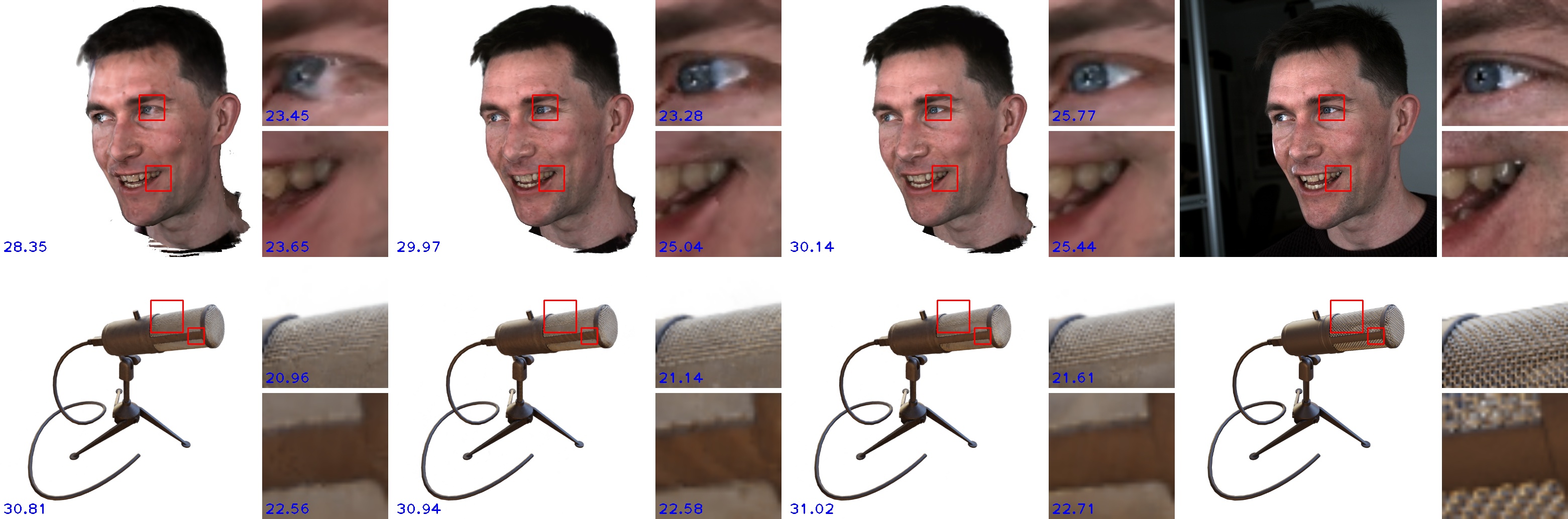}
  \begin{subfigure}{.24\textwidth}
  \centering
  \caption{2 MLPs}
  \end{subfigure}
  \begin{subfigure}{.24\textwidth}
  \centering
  \caption{1 MLP, NeRF HS}
  \end{subfigure}
  \begin{subfigure}{.24\textwidth}
  \centering
  \caption{1 MLP, our HS}
  \end{subfigure}
  \begin{subfigure}{.24\textwidth}
  \centering
  \caption{Ground truth}
  \end{subfigure}
  \caption{\ready{} Novel-view renders using different sampling strategies with a fixed total number of network evaluations. The default strategy of using two Multi-Layer Perceptrons (MLPs) ((a), first column), does not use the samples in an efficient way, missing thin structures like the bottom teeth. Naively using a single network with the default hierarchical sampling (HS) of NeRF ((b), second column) can improve sharpness, but leads to ring-like artefacts, for example, on the lip or on the brown strip in the mic scene. Using one network with our proposed hierarchical sampling ((c), third column) improves the sharpness on the eyes, teeth and fine mic structure while avoiding the artefacts. The number in blue in the bottom-left corner of each image is the Peak Signal-to-Noise Ratio (PSNR) of the render or its crop.}
  \label{fig:05_experiments/ablations/importance_sampling}
\end{figure*}

\ready{} \textbf{Sampling}. To evaluate the impact of our improved sampling strategy, we use NeRF Synthetic datasets~\cite{nerf} and static versions of our face datasets (30 images from a single timestamp in each of the four faces). We train models for three different scenarios each with the same number total network evaluations per ray: (1) two-network setting with 64 samples for the coarse network and 64 samples for the fine network, (2) one-network setting with 128 coarse samples and 64 fine samples with NeRF's hierarchical sampling, and (3) one-network setting with our proposed hierarchical sampling.  Figure~\ref{fig:05_experiments/ablations/importance_sampling} compares qualitatively the three sampling strategies for novel-view rendering. Our sampling strategy results in sharper images than the two-MLP setting because we double the number of coarse samples and it removes the banding artefacts of a single MLP with NeRF's hierarchical sampling because it does not underestimate boundaries. This observation is supported by quantitative results in Table~\ref{tab:05_experiments/ablations/importance_sampling}.

\begin{figure}
  \centering
  \begin{subfigure}{0.45\textwidth}
  \centering
  \includegraphics[height=2.1cm]{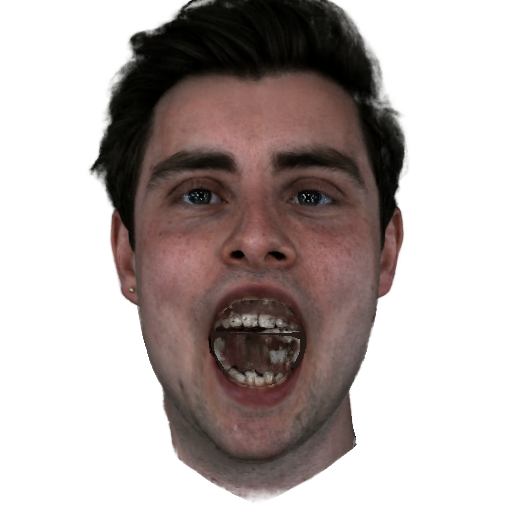}  
  \includegraphics[height=2.5cm]{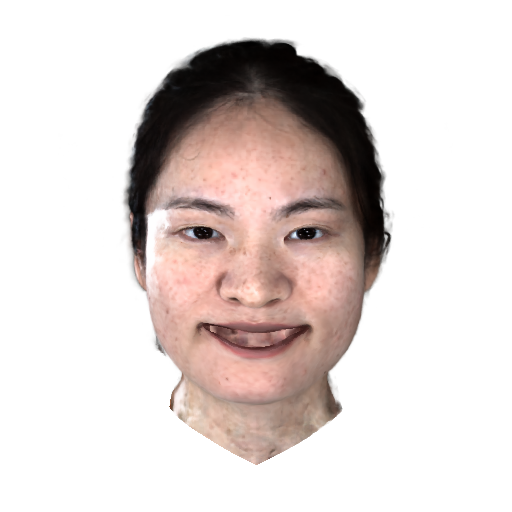}  
  \includegraphics[height=2.5cm]{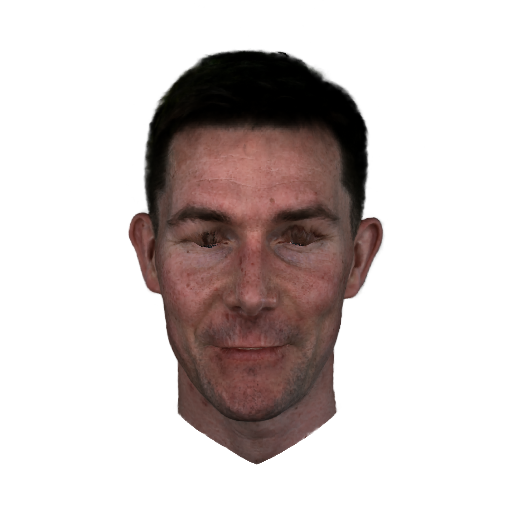}
  \caption{Failure because mouth interior and closed eyelids unseen in training.}
  \label{fig:failure_unobserved_regions}
\end{subfigure}
\\
  \begin{subfigure}{0.45\textwidth}
  \centering
  \includegraphics[height=2.1cm]{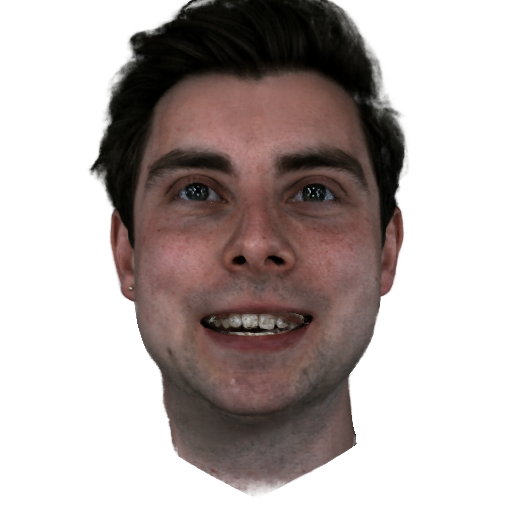}  
   \includegraphics[height=2.5cm]{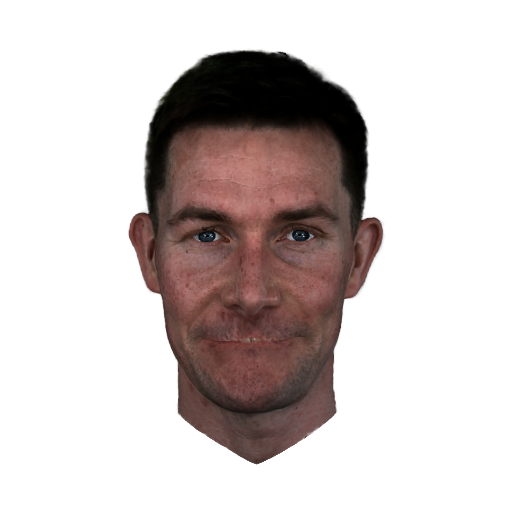}
    \includegraphics[height=2.5cm]{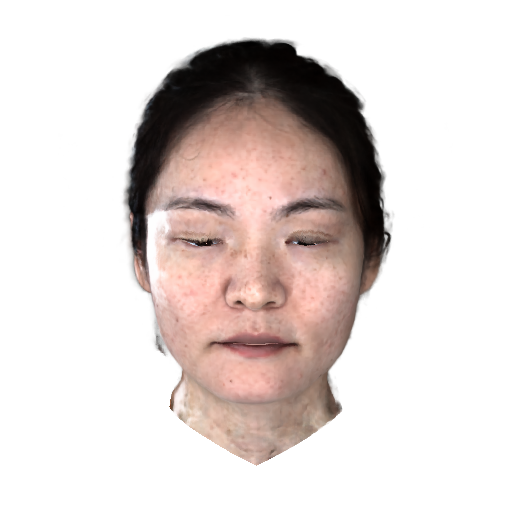}  
  \caption{Failures because lips and eyelids are poorly fit in training.}
  \label{fig:failure_fit}
\end{subfigure}
\\
  \begin{subfigure}{0.45\textwidth}
  \centering
  \includegraphics[height=2.5cm]{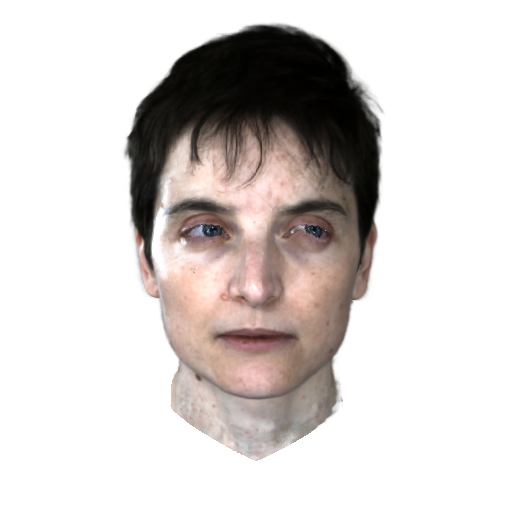} 
    \includegraphics[height=2.5cm]{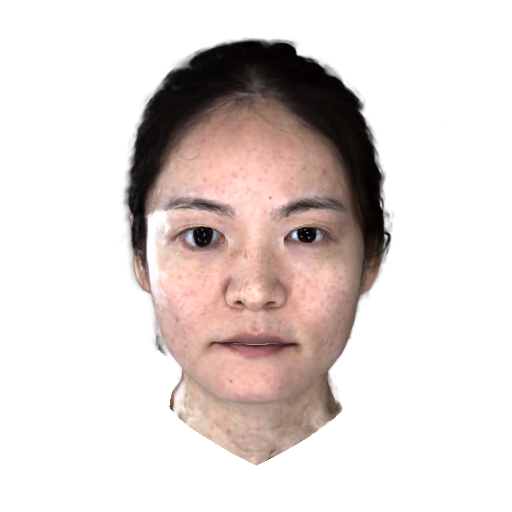} 
     \includegraphics[height=2.5cm]{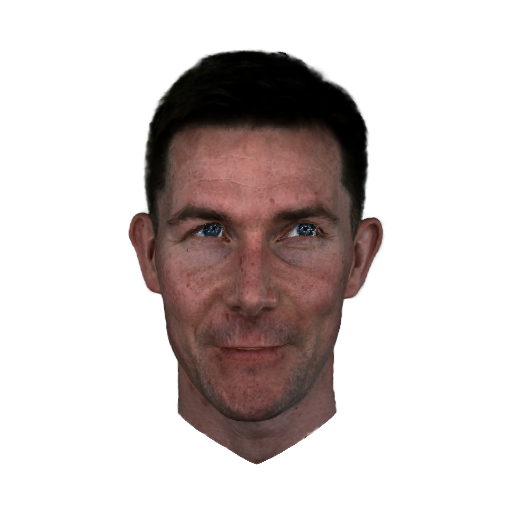} 
  \caption{Failure because tetrahedra on eyeballs and eyelids share deformation.}
  \label{fig:failure_eyes}
\end{subfigure}
  \caption{\ready{} \textbf{Limitations} our method a) cannot accurately generate the appearance of regions unseen in training \eg{} the mouth interior, b) requires good alignment of the fitted face model to the training frame(s) to ensure that face regions deform as intended by the face model, and c) constrains eyeballs and eyelids to move jointly which can create unrealistic stretching of the eye region for extreme gaze or lid motions.}
  \label{fig:failure_modes}
\end{figure}

\section{Limitations and future work}
\label{section:limitations_future}
\ready{}

Our method has three main limitations. First, the interpolation in our volumetric meshes is linear, and so very large or intricate deformations require high resolution geometry. This can be addressed by using techniques such as \cite{meanValueCoordinatesPaper} at increased computational cost. Second, our acceleration requires hardware ray tracing for best performance. While ubiquitous in high-end desktop hardware, mobile devices still mostly lack such implementations. Finally, our method assumes no self-intersections of the geometry, similar to what tetrahedral FEM models require. This means care has to be taken when constructing the geometry to avoid visual artefacts.

\textbf{\methodname{} applied to objects.} For the experiments with generic objects, we defined the topology and the position of vertices of the tetrahedral mesh in a partly manual process.
Future work includes using machine learning to automatically predict the position of the vertices of the tetrahedra~ \cite{Yifan:NeuralCage:2020} and support for pose dependent appearance. 

\textbf{\methodname{} applied to avatars.} Our model has three failure modes when applied to faces: unseen regions, poor fit of training frame, shared deformation of eyeballs and eye region.
Our model cannot generate regions unseen in training: expressions that expose all the teeth, lip interior, or complete eyelids results in artefacts or unrealistic stretching of the region, as shown in Figure \ref{fig:failure_unobserved_regions}. Our model relies on an accurate fit of the training frame to deform the samples into a canonical pose. This is specially important for regions with large motion, like the lips and eyes, as errors in the fit result in wrong deformation of the samples and a canonical pose where multiple face regions collide. For instance in \ref{fig:failure_fit}, a poor fit around the lips deforms the teeth with the lips and learns a canonical pose where teeth are inpainted on the lip and lips inpainted on teeth. A poor fit around the eyelids can result in avatars that do not to fully close the eyes because some of the eyeball is inpainted into the lid. Finally, in our model, the motion of the eyeballs and the eye region are all controlled by the same set of tetrahedra. As a result, gaze motion in the eyeballs leaks to the boundary between eyes and lids. Similarly, motion of the eyelids can leak into the eyeballs and stretch them. Figure \ref{fig:failure_eyes} shows examples of both failure cases. Future work includes leveraging a generative photo-real head model~\cite{avaps,morf} and to move away from the multi-view single frame regime to the monocular multi-frame regime, which is effectively the setup used by~\cite{avaps,neural_head_avatars,nerfies,Gafni_2021_CVPR}. 
\section{Conclusions}
\label{section:conclusion}

\ready{}
In this work we introduced \methodname{}, the first volumetric system capable of generating photo-realistic dynamic content that can be enrolled from a limited amount of data, can be controlled using established control mechanisms like physics-based simulation and blendshapes, can render in real-time, and can support both objects and avatars. We presented applications of our system to dynamic objects and avatars that experimentally improve on the state of the art and show the versatility of our framework. We believe these promising results demonstrate the potential of \methodname{} for developing high quality immersive experiences in the Metaverse, AR, and VR.

%
%
%
%

\bibliographystyle{ACM-Reference-Format}
\bibliography{sample-bibliography}


\end{document}